\documentclass[twocolumn]{aastex62}

\shorttitle{SFRS: 3 - 33\,GHz Imaging of Star-Forming Regions}
\shortauthors{Linden et al.}

\begin{document}

\title{The Star Formation in Radio Survey: 3 -- 33\,GHz Imaging of Nearby Galaxy Nuclei and Extranuclear Star-forming Regions}

\author{S.T.\,Linden}
\affiliation{Department of Astronomy, University of Virginia, 530 McCormick Road, Charlottesville, VA 22904, USA}

\author{E.J.\,Murphy}
\affiliation{National Radio Astronomy Observatory, 520 Edgemont Road, Charlottesville, VA 22903, USA}

\author{D.\,Dong}
\affiliation{California Institute of Technology, MC 100-22, Pasadena, CA 91125, USA}

\author{E.\,Momjian}
\affiliation{National Radio Astronomy Observatory, P.O. Box O, 1003 Lopezville Road, Socorro, NM 87801, USA}

\author{R. C.\,Kennicutt Jr.}
\affiliation{Institute of Astronomy, University of Cambridge, Madingley Road, Cambridge CB3 0HA, UK}

\author{D.S.\,Meier}
\affiliation{New Mexico Institute of Mining and Technology, 801 Leroy Place, Socorro, NM 87801, USA}

\author{E.\,Schinnerer}
\affiliation{Max Planck Institut f\"{u}r Astronomie, Königstuhl 17, Heidelberg D-69117, Germany}

\author{J.L.\,Turner}
\affiliation{Department of Physics and Astronomy, UCLA, Los Angeles, CA 90095, USA}

\date{\today}

\begin{abstract}
%A fundamental goal of this survey is to map the current star formation activity on $\sim$100pc scales within a large, heterogeneous, sample of nearby ($d_{L} \leq 30$ Mpc) galaxies without the additional complications of spatially-varying dust extinction.
We present 3, 15, and 33\,GHz imaging towards galaxy nuclei and extranuclear star-forming regions using the Karl G. Jansky Very Large Array as part of the Star Formation in Radio Survey.~With $3-33$\,GHz radio spectra, we measured the spectral indices and corresponding thermal (free-free) emission fractions for a sample of 335 discrete regions having significant detections in at least two radio bands.~After removing 14 likely background galaxies, we find that the median thermal fraction at 33\,GHz is $92 \pm 0.8\%$ with a median absolute deviation of $11\%$, when a two-component power-law model is adopted to fit the radio spectrum. Limiting the sample to 238 sources that are confidently identified as star-forming regions, and not affected by potential AGN contamination (i.e., having galactocentric radii $r_{\rm G} \geq 250$\,pc), results in a median thermal fraction of $93 \pm 0.8$\% with a median absolute deviation of $10\%$. We further measure the thermal fraction at 33\,GHz for 163 regions identified at 7\arcsec~resolution to be $94 \pm 0.8$\% with a median absolute deviation of $8\%$. Together, these results confirm that free-free emission dominates the radio spectra of star-forming regions on scales up to $\sim$500\,pc in normal star-forming galaxies. We additionally find a factor of $\sim$1.6 increase in the scatter of the measured spectral index and thermal fraction distributions as a function of decreasing galactocentric radius. This trend is likely reflective of the continuous star-formation activity occurring in the galaxy centers, resulting a larger contribution of diffuse nonthermal emission relative to star-forming regions in the disk.
%~Finally, we identify a sample of 33 anomalous microwave emission (AME) candidates, whose rising $3-33$\,GHz spectra may be due to an additional emission component at $\sim 20-50$\,GHz. %Follow-up observations at high ($\geq 40$\,GHz) frequencies will be necessary to confirm these sources as discrete regions of extragalactic AME.

\end{abstract}

\keywords{galaxies: nuclei - H{\sc ii} regions - radio continuum: general - stars: formation}

%IMPORTANT info about tables: 6 regions in 7\arcsec~Table 6 have 0.00 as an error for Ha measurement. The sigfigs have been increased to 3 for these cases.
%~25 regions in native res table with 0.00 errors in f_th and spx. The sigfigs have been increased to 3 for ALL fth and spx measurements at native-resolution in Table 4.

\section{Introduction}

The radio spectra of star-forming galaxies, typically characterized as a power law ($S_\nu \propto \nu^{\alpha}$), encode information about the thermal and non-thermal energetic processes which power them. Both thermal (Bremsstrahlung) and non-thermal (synchrotron) emission are associated with massive ($\geq 8 M_{\odot}$) star formation, underlying the basis for the well-known far-infrared (FIR: 42-122$\mu$m)-radio correlation \citep{de85,gxh85,jc92, efb03}. FIR emission arises from the absorption and re-radiation of UV and optical photons that heat dust grains surrounding massive star-forming regions. The O and B stars in such regions, with lifetimes of $\leq 10$ Myr, produce ionizing (Lyman continuum) radiation whose strength is directly proportional to the amount of free-free emission.~These same massive stars end their lives as core-collapse supernovae, whose remnants accelerate cosmic ray (CR) electrons/positrons that produce the diffuse non-thermal synchrotron emission observed in star-forming galaxies \citep{jc92,kk95,ejm06a,lt10,ltq10}.

However, the connection between the non-thermal synchrotron emission and the current star-formation rate (SFR) of a galaxy is far less direct relative to thermal free-free emission. Variations in the generation and propagation of CRs through the interstellar medium (ISM) can affect the observed low-frequency emission surrounding star-forming regions. Despite the complexity in interpreting this non-thermal emission from galaxies, several empirical \citep{efb03, kk95,ejm06a, hv14, fat17} and theoretical \citep{jc92, ejm11b} calibrations for the star-formation rate (SFR) exist in the literature.~These studies demonstrate that at frequencies low enough (typically $\sim 1$\,GHz) the emission is dominated by the non-thermal, steep spectrum component ($\alpha^{\rm NT} \sim -0.8$), and at frequencies high enough ($\sim 30$\,GHz) the emission becomes dominated by thermal emission ($\alpha^{\rm T} \sim -0.1$). Hence, radio observations can serve as an excellent, extinction-free, diagnostic for the current SFR within nearby galaxies.

This was the motivation for initiating the Star Formation in Radio Survey (SFRS), which began as a 33\,GHz imaging campaign with the Green Bank Telescope (GBT) to study 103 galaxy nuclei and extranuclear star-forming complexes at a matched resolution of $25\arcsec$.~In the initial investigation, \citet{ejm12b} used the Westerbork Synthesis Radio Telescope (WSRT) in combination with the GBT to construct 1.7-to-33\,GHz radio spectra for 53 galaxy nuclei and extranuclear star-forming regions on $\sim$kpc scales.~They found evidence that the measured thermal fraction at 33\,GHz varied significantly for star-forming regions observed at different physical resolution due to the range in galaxy distance. Photometric apertures larger than $\sim 1$kpc were observed to have thermal fractions as low as $40-50\%$, whereas regions measured with apertures $\leq$1kpc appeared to be heavily dominated by free-free emission, with thermal fractions as high as $\sim90\%$.~However, without high-resolution maps of both free-free and non-thermal emission from individual H{\sc ii} regions within these larger complexes, it is difficult to determine the physical nature of these trends within nearby galaxies.

This study served as the foundation for extending the SFRS into a multi-frequency Karl G. Jansky Very Large Array (VLA) campaign to image hundreds of star-forming regions in 50 galaxies taken from the \textit{Spitzer} Infrared Nearby Galaxies Survey \citep[SINGS:][]{rck03} and the Key Insights on Nearby Galaxies:~a Far-Infrared Survey \citep[KINGFISH:][]{kf11}.~The results from our 33\,GHz observations, along with corresponding H$\alpha$ and {\it Spitzer}/MIPS 24$\mu$m photometry were recently presented in \citet{ejm18a}, hereafter M18a, and explored the H$\alpha$-to-33\,GHz and 24$\mu$m-to-33\,GHz flux density ratios of star-forming regions as a function of galactocentric radius and physical resolution.~An outlier of these distributions, NGC\,4725\,B, was later followed up with higher-frequency observations (Q-band: $\sim$ 44\,GHz) in order to confirm this region as the second known source of extragalactic anomalous microwave emission \citep[AME:][]{ejm18b}.~Building on this analysis, we have also obtained 3 and 15\,GHz imaging for the SFRS, allowing us to map the full radio spectrum of each star-forming region at a matched-resolution of $\sim 2\arcsec$. In this paper, we focus our presentation on the results associated with the radio spectral indices and corresponding free-free emission fractions for the entire sample. 

%By having frequency coverage below 10\,GHz our observations are sensitive to how the environments of star-forming galaxies affect the CR propagation and ionized gas physics of individual H{\sc ii} regions. In particular, low-frequency radio spectra are useful in constructing studying cosmic ray propagation length, which can be used to constrain the strength of the magnetic fields within star-forming galaxies (e.g., Torres 2004; Lacki et al. 2010; Tabatabei et al. 2017). Further, we are able to quantify the level at which the radio spectra of typical H{\sc ii} regions are affected by free-free absorption, which may be important in the centers of highly star-forming and starburst galaxies (e.g. Barcos-Munoz et al. 2015). For individual (ultra-compact) H{\sc ii} regions, the turnover frequency can be as high as $\sim20$\,GHz (Murphy et al. 2010). 
%Consequently, two fundamental goals of this survey are to map star formation activity through ionizing photon rates and characterize cosmic ray propagation for a large, heterogeneous, sample of star-forming regions in nearby galaxies, without the additional complications of extinction and contamination, often found with optical and near-infrared (NIR: [1-5$\mu$m]) studies.

The paper is organized as follows: In \S2 we describe our sample selection, data reduction, and imaging procedure for the 3 - 33\,GHz VLA data. In \S3 we describe the ancillary data products included in this study as well as the analysis procedures used. Our results are presented in \S4, and discussed in \S5. Our main conclusions are summarized in \S6.~In the Appendix we additionally provide ancillary photometry from the Galaxy Evolution Explorer (GALEX), \textit{Spitzer}, and ground-based H$\alpha$ observations at a matched resolution of 7\arcsec~that are not used in the present analysis. Throughout the paper we report the median absolute deviations rather than standard deviations, as this statistic is more resilient against outliers in a data set.

\section{Sample and Data Analysis}

In this section we describe the sample selection, and present the VLA observations along with the data reduction and imaging procedures.

\begin{deluxetable*}{l|cccccc}
\tablecaption{Galaxy Properties \label{tbl-1}}
\tabletypesize{\scriptsize}
\tablewidth{0pt}
\tablehead{
\colhead{Galaxy}  & \colhead{Type\tablenotemark{a}} & \colhead{Dist.\tablenotemark{b}} & \colhead{Nuc. Type\tablenotemark{c}}  & \colhead{$D_{25}$\tablenotemark{a}} & \colhead{$i$} & \colhead{P.A.\tablenotemark{a}} \\
\colhead{} & \colhead{} & \colhead{(Mpc)} & \colhead{} & \colhead{(arcmin)} & \colhead{($\degr$)} & \colhead{($\degr$)} 
}
\startdata
   NGC\,0337     &      SBd  &19.3  &                SF  &  $2.9 \times 1.8$  & 52  &               130\\
   NGC\,0628     &      SAc  & 7.2  &           \nodata  & $10.5 \times 9.5$  & 25  &                25\\
   NGC\,0855     &        E  &9.73  &                SF  &  $2.6 \times 1.0$  & 70  &   67\rlap{$^{d}$}\\
   NGC\,0925     &     SABd  &9.12  &                SF  & $10.5 \times 5.9$  & 57  &               102\\
   NGC\,1097     &      SBb  &14.2  &               AGN  &  $9.3 \times 6.3$  & 48  &               130\\
   NGC\,1266     &      SB0  &30.6  &               AGN  &  $1.5 \times 1.0$  & 49  &  108\rlap{$^{d}$}\\
   NGC\,1377     &       S0  &24.6  &           \nodata  &  $1.8 \times 0.9$  & 61  &                92\\
    IC\,0342       &    SABcd  &3.28  &             SF(*)  & $21.4 \times 20.$  & 21  &  153\rlap{$^{d}$}\\
   NGC\,1482     &      SA0  &22.6  &                SF  &  $2.5 \times 1.4$  & 57  &               103\\
   NGC\,2146     &     Sbab  &17.2  &             SF(*)  &  $6.0 \times 3.4$  & 56  &                57\\
   NGC\,2403     &    SABcd  &3.22  &             SF(*)  &$21.9 \times 12.3$  & 57  &               128\\
Holmberg\,II     &       Im  &3.05  &           \nodata  &  $7.9 \times 6.3$  & 37  &                16\\
   NGC\,2798     &      SBa  &25.8  &            SF/AGN  &  $2.6 \times 1.0$  & 70  &               160\\
   NGC\,2841     &      SAb  &14.1  &               AGN  &  $8.1 \times 3.5$  & 66  &               147\\
   NGC\,2976     &      SAc  &3.55  &                SF  &  $5.9 \times 2.7$  & 64  &               143\\
   NGC\,3049     &     SBab  &19.2  &                SF  &  $2.2 \times 1.4$  & 51  &                25\\
   NGC\,3077     &    I0pec  &3.83  &             SF(*)  &  $5.4 \times 4.5$  & 34  &                45\\
   NGC\,3190     &     SAap  &19.3  &            AGN(*)  &  $4.4 \times 1.5$  & 73  &               125\\
   NGC\,3184     &    SABcd  &11.7  &                SF  &  $7.4 \times 6.9$  & 21  &               135\\
   NGC\,3198     &      SBc  &14.1  &                SF  &  $8.5 \times 3.3$  & 68  &                35\\
    IC\,2574      &     SABm  &3.79  &             SF(*)  & $13.2 \times 5.4$  & 67  &                50\\
   NGC\,3265     &        E  &19.6  &                SF  &  $1.3 \times 1.0$  & 39  &                73\\
   NGC\,3351     &      SBb  &9.33  &                SF  &  $7.4 \times 5.0$  & 48  &                13\\
   NGC\,3521     &    SABbc  &11.2  &         SF/AGN(*)  & $11.0 \times 5.1$  & 63  &               163\\
   NGC\,3621     &      SAd  &6.55  &               AGN  & $12.3 \times 7.1$  & 55  &               159\\
   NGC\,3627     &     SABb  &9.38  &               AGN  &  $9.1 \times 4.2$  & 64  &               173\\
   NGC\,3773     &      SA0  &12.4  &                SF  &  $1.2 \times 1.0$  & 33  &               165\\
   NGC\,3938     &      SAc  &17.9  &             SF(*)  &  $5.4 \times 4.9$  & 25  &   29\rlap{$^{d}$}\\
   NGC\,4254     &      SAc  &14.4  &            SF/AGN  &  $5.4 \times 4.7$  & 30  &   24\rlap{$^{d}$}\\
   NGC\,4321     &    SABbc  &14.3  &               AGN  &  $7.4 \times 6.3$  & 32  &                30\\
   NGC\,4536     &    SABbc  &14.5  &            SF/AGN  &  $7.6 \times 3.2$  & 66  &               130\\
   NGC\,4559     &    SABcd  &6.98  &                SF  & $10.7 \times 4.4$  & 67  &               150\\
   NGC\,4569     &    SABab  &9.86  &               AGN  &  $9.5 \times 4.4$  & 64  &                23\\
   NGC\,4579     &     SABb  &16.4  &               AGN  &  $5.9 \times 4.7$  & 37  &                95\\
   NGC\,4594     &      SAa  &9.08  &               AGN  &  $8.7 \times 3.5$  & 69  &                90\\
   NGC\,4625     &    SABmp  & 9.3  &                SF  &  $2.2 \times 1.9$  & 31  &   28\rlap{$^{d}$}\\
   NGC\,4631     &      SBd  &7.62  &             SF(*)  & $15.5 \times 2.7$  & 83  &                86\\
   NGC\,4725     &    SABab  &11.9  &               AGN  & $10.7 \times 7.6$  & 45  &                35\\
   NGC\,4736     &     SAab  &4.66  &            AGN(*)  & $11.2 \times 9.1$  & 35  &               105\\
   NGC\,4826     &     SAab  &5.27  &               AGN  & $10.0 \times 5.4$  & 59  &               115\\
   NGC\,5055     &     SAbc  &7.94  &               AGN  & $12.6 \times 7.2$  & 56  &               105\\
   NGC\,5194     &   SABbcp  &7.62  &               AGN  & $11.2 \times 6.9$  & 53  &               163\\
   NGC\,5398     &     SBdm  &7.66  &           \nodata  &  $2.8 \times 1.7$  & 53  &               172\\
   NGC\,5457     &    SABcd  & 6.7  &             SF(*)  & $28.8 \times 26.$  & 26  &   29\rlap{$^{d}$}\\
   NGC\,5474     &     SAcd  & 6.8  &             SF(*)  &  $4.8 \times 4.3$  & 27  &   98\rlap{$^{d}$}\\
   NGC\,5713     &   SABbcp  &21.4  &                SF  &  $2.8 \times 2.5$  & 27  &                10\\
   NGC\,5866     &       S0  &15.3  &               AGN  &  $4.7 \times 1.9$  & 69  &               128\\
   NGC\,6946     &    SABcd  & 6.8  &                SF  & $11.5 \times 9.8$  & 32  &   53\rlap{$^{d}$}\\
   NGC\,7331     &      SAb  &14.5  &               AGN  & $10.5 \times 3.7$  & 72  &               171\\
   NGC\,7793     &      SAd  &3.91  &                SF  &  $9.3 \times 6.3$  & 48  &                98  
\enddata
\tablenotetext{a}{Morphological types, diameters, and position angles were taken from the Third Reference Catalog of Bright Galaxies \citep[RC3:][]{dev91}.}%the NASA/IPAC Extragalactic Database (NED; http://nedwww.ipac.caltech.edu).}
\tablenotetext{b}{Redshift-independent distance taken from the list compiled by \citet{kf11}, except for the two non-KINGFISH galaxies NGC\,5194 \citep{rc02} and NGC\,2403 \citep{hkp01}.}
\tablenotetext{c}{Nuclear type based on optical spectroscopy: SF$\,=\,$ Star-Forming; AGN$\,=\,$ Non-thermal emission as given in Table~5 of \citet{jm10} or (*) Table~4 of \citet{lh87}.}
\tablenotetext{d}{Position angle taken from \citet{tj03}.}
\end{deluxetable*}

\subsection{Sample Selection}

The SFRS sample includes targeted observations from 56 nearby galaxies ($d_{L} < 30$ Mpc) in the SINGS and KINGFISH legacy programs (Table 1). All nuclear and extranuclear star-forming regions were chosen to have mid-infrared spectral mappings carried out by the IRS instrument on board \textit{Spitzer}, and \textit{Herschel}/PACS far-infrared spectral mappings, for a combination of the principal atomic ISM cooling lines ([OI] 63$\mu$m, [OIII] 88$\mu$m, [NII] 122, 205$\mu$m, and [CII] 158$\mu$m). NGC\,5194 and NGC\,2403 are exceptions; these galaxies were part of the SINGS sample, but are not formally included in KINGFISH. They were observed with \textit{Herschel} as part of the Very Nearby Galaxy Survey (VNGS; PI: C. Wilson). Similarly, there are additional KINGFISH galaxies that were not part of SINGS, but have existing \textit{Spitzer} data: NGC\,5457 (M101), IC\,342, NGC\,3077, and NGC\,2146.

The SINGS and KINGFISH galaxies are fully representative of the integrated properties and ISM conditions found in the local Universe, spanning the full range in morphological types, as well as a factor of 100 in IR luminosity ($L_{\rm IR}$:\,8-1000$\mu$m), global IR/optical flux ratio, and the star formation rate. Similarly, the spectroscopically targeted star-forming regions included in the SFRS cover the full range of physical conditions found in nearby galaxies, including the extinction-corrected production rate of ionizing photons $Q(H_{0})$, metallicity, visual extinction, radiation field intensity, and ionizing stellar temperature. 

The full sample over the entire sky consists of 118 star-forming complexes (56 nuclei and 62 extranuclear regions), 112 of which (50 nuclei and 62 extranuclear regions; see Tables 2 and 3, respectively) are observable with the VLA (i.e., having $\delta > -35\degr$). The coordinates given in both tables list the pointing center for the VLA observations (see Section 2.2), which correspond to the centers of the \textit{Spitzer} mid-infrared and \textit{Herschel} far-infrared spectral line maps. 
Morphologies, adopted distances, optically-defined nuclear types, diameters ($D_{25}$), inclinations (\textit{i}), and position angles (P.A.) for each source are given in Table 1 and described in detail in M18a.

\subsection{VLA Observations and Data Reduction}

The observational set-up and reduction procedure for the Ka-band ($29 - 37$\,GHz) data (11B-032,13A-129) is described in detail in M18a. Observations in the S-band ($2-4$\,GHz) were taken during the 2013 VLA B-configuration cycle (13B-215), and utilized the 8-bit sampler. Observations in the Ku-band ($12-18$\,GHz) were taken November 2014 in the C-configuration (13B-215) using the 3-bit samplers. Both sets of observations utilized the full available bandwidth of the respective receivers. Given the large range in brightness among our targeted regions, we varied the time spent on-source by estimating the expected $3-15$\,GHz flux density using the \textit{Spitzer}/MIPS 24\,$\mu$m maps. The median integration time for regions in our sample was $\sim$ 10 minutes at both frequencies. The choice of array configurations were made to match the angular resolution (i.e., FWHM of the synthesized beam $\sim 2\arcsec$) of the observations at each band. This allows us to probe the same spatial scales across the full $3 - 33$\,GHz frequency range, and ensures that any differences in the measured spectral index of individual star-forming regions is due to physical variation in the region being measured, and not due to resolving out more emission at higher frequencies.

The standard VLA flux density calibrators 3C48, 3C286, and 3C147 were used, and the data reduction procedures presented in M18a are repeated for our present analysis, and briefly described here. To reduce the VLA data, we used the Common Astronomy Software Applications \citep[CASA;][]{casa} versions 4.6.0 and 4.7.0, and followed standard calibration and flagging procedures, including the utilization of the VLA calibration pipeline. We further inspected the visibilities and calibration tables for evidence of bad antennas, frequency ranges, and time ranges, flagging correspondingly. We also flagged any instances of radio frequency interference (RFI). Importantly, the fractional bandwidth of our observations lost to RFI flagging is negligible relative to the full bandwidth of the receivers. After flagging, we re-ran the pipeline, and repeated this process until all poorly-calibrated data were removed. For all delay and bandpass tables applied on-the-fly, we used the default nearest-neighbor interpolation. For complex gain and flux density scale tables, we used a linear interpolation. 

\begin{deluxetable*}{l|cc|ccc|ccc}
\tablecaption{Nuclear Source Positions and Imaging Characteristics \label{tbl-2}}
\tabletypesize{\scriptsize}
\tablewidth{0pt}
\tablehead{
\colhead{} & \colhead{} & \colhead{} & \multicolumn{3}{|c|}{3\,GHz} & \multicolumn{3}{c}{15\,GHz} \\
\colhead{Galaxy} & \colhead{R.A.} & \colhead{Decl.} &  \colhead{Synthesized}& \colhead{$\sigma$}& \colhead{$\sigma_{T_{\rm b}}$} & \colhead{Synthesized}& \colhead{$\sigma$}& \colhead{$\sigma_{T_{\rm b}}$}\\
\colhead{}  & \colhead{(J2000)} & \colhead{(J2000)} & \colhead{Beam}& \colhead{($\mu$Jy\,bm$^{-1}$)}& \colhead{(mK)} & \colhead{Beam}& \colhead{($\mu$Jy\,bm$^{-1}$)}& \colhead{(mK)}
}
\startdata
           NGC\,0337   &   $00~59~50.3$  &    $-07~34~44$  &  $2\farcs43 \times 1\farcs74$  &18.7  & 597.32  &  $2\farcs00 \times 1\farcs13$  & 9.1  & 21.71\\
           NGC\,0628   &   $01~36~41.7$  &    $+15~46~59$  &  $1\farcs96 \times 1\farcs78$  &14.0  & 543.15  &  $1\farcs55 \times 1\farcs19$  & 9.6  & 28.14\\
           NGC\,0855   &   $02~14~03.7$  &    $+27~52~38$  &  $1\farcs80 \times 1\farcs62$  &13.2  & 613.61  &  $1\farcs50 \times 1\farcs25$  & 8.6  & 24.93\\
           NGC\,0925   &   $02~27~17.0$  &    $+33~34~43$  &  $1\farcs80 \times 1\farcs59$  &13.0  & 611.30  &  $1\farcs47 \times 1\farcs23$  & 9.1  & 27.04\\
           NGC\,1097   &   $02~46~19.1$  &    $-30~16~28$  &  $5\farcs76 \times 1\farcs80$  &44.9  & 586.45  &  $3\farcs81 \times 0\farcs99$  &16.9  & 24.11\\
           NGC\,1266   &   $03~16~00.8$  &    $-02~25~38$  &  $2\farcs20 \times 1\farcs78$  &14.0  & 485.33  &  $1\farcs80 \times 1\farcs17$  &11.3  & 28.86\\
           NGC\,1377   &   $03~36~38.9$  &    $-20~54~06$  &  $3\farcs56 \times 1\farcs71$  &17.4  & 385.40  &  $2\farcs61 \times 1\farcs09$  &11.6  & 21.93\\
            IC\,0342   &   $03~46~48.5$  &    $+68~05~46$  &  $2\farcs23 \times 1\farcs76$  &41.6  &1430.61  &  $1\farcs72 \times 1\farcs13$  &11.4  & 31.44\\
           NGC\,1482   &   $03~54~39.5$  &    $-20~30~07$  &  $3\farcs25 \times 1\farcs67$  &16.8  & 419.18  &  $2\farcs42 \times 1\farcs33$  &14.1  & 23.56\\
           NGC\,2146   &   $06~18~37.7$  &    $+78~21~25$  &  $2\farcs55 \times 1\farcs55$  &35.5  &1214.57  &  $1\farcs93 \times 0\farcs94$  &14.9  & 44.68\\
           NGC\,2403   &   $07~36~50.0$  &    $+65~36~04$  &  $2\farcs21 \times 1\farcs53$  &13.8  & 551.32  &  $1\farcs88 \times 1\farcs15$  & 9.8  & 24.41\\
        Holmberg\,II   &   $08~19~13.3$  &    $+70~43~08$  &  $2\farcs59 \times 1\farcs67$  &14.9  & 465.55  &  $1\farcs97 \times 1\farcs15$  & 9.4  & 22.41\\
           NGC\,2798   &   $09~17~22.8$  &    $+41~59~58$  &  $2\farcs13 \times 1\farcs78$  &14.6  & 520.66  &  $1\farcs58 \times 1\farcs38$  &11.3  & 27.83\\
           NGC\,2841   &   $09~22~02.7$  &    $+50~58~36$  &  $2\farcs11 \times 1\farcs62$  &13.3  & 526.88  &  $1\farcs59 \times 1\farcs23$  &18.1  & 50.13\\
           NGC\,2976   &   $09~47~15.3$  &    $+67~55~00$  &  $2\farcs83 \times 1\farcs66$  &14.4  & 413.49  &  $2\farcs02 \times 1\farcs13$  & 9.7  & 23.08\\
           NGC\,3049   &   $09~54~49.6$  &    $+09~16~17$  &  $2\farcs14 \times 1\farcs84$  &14.9  & 510.77  &  $2\farcs07 \times 1\farcs13$  &12.2  & 27.98\\
           NGC\,3077   &   $10~03~19.1$  &    $+68~44~02$  &  $2\farcs90 \times 1\farcs67$  &14.4  & 402.59  &  $1\farcs98 \times 1\farcs12$  &10.2  & 24.82\\
           NGC\,3190   &   $10~18~05.6$  &    $+21~49~55$  &  $2\farcs12 \times 1\farcs81$  &13.5  & 475.29  &  $1\farcs61 \times 1\farcs26$  & 8.5  & 22.63\\
           NGC\,3184   &   $10~18~16.7$  &    $+41~25~27$  &  $2\farcs42 \times 1\farcs76$  &14.0  & 444.09  &  $1\farcs29 \times 1\farcs18$  & 8.5  & 30.49\\
           NGC\,3198   &   $10~19~54.9$  &    $+45~32~59$  &  $2\farcs25 \times 1\farcs66$  &14.4  & 521.10  &  $1\farcs34 \times 1\farcs17$  & 8.5  & 29.06\\
            IC\,2574   &   $10~28~48.4$  &    $+68~28~02$  &  $2\farcs96 \times 1\farcs64$  &14.2  & 393.53  &  $2\farcs03 \times 1\farcs14$  & 8.8  & 20.73\\
           NGC\,3265   &   $10~31~06.7$  &    $+28~47~48$  &  $2\farcs23 \times 1\farcs78$  &14.9  & 505.30  &  $1\farcs44 \times 1\farcs25$  & 7.8  & 23.42\\
           NGC\,3351   &   $10~43~57.8$  &    $+11~42~14$  &  $2\farcs00 \times 1\farcs75$  &18.4  & 708.89  &  $1\farcs95 \times 1\farcs67$  &15.3  & 25.34\\
           NGC\,3521   &   $11~05~48.9$  &    $-00~02~06$  &  $3\farcs07 \times 1\farcs95$  &19.9  & 448.93  &  $1\farcs88 \times 1\farcs19$  &16.4  & 39.79\\
           NGC\,3621   &   $11~18~16.0$  &    $-32~48~42$  &  $4\farcs76 \times 1\farcs53$  &15.7  & 291.11  &  $4\farcs17 \times 1\farcs08$  & 8.5  & 10.16\\
           NGC\,3627   &   $11~20~15.0$  &    $+12~59~30$  &  $1\farcs95 \times 1\farcs77$  &15.7  & 616.48  &  $2\farcs36 \times 1\farcs97$  &14.0  & 16.23\\
           NGC\,3773   &   $11~38~13.0$  &    $+12~06~45$  &  $1\farcs95 \times 1\farcs76$  &13.1  & 518.26  &  $1\farcs53 \times 1\farcs14$  & 9.8  & 30.21\\
           NGC\,3938   &   $11~52~49.5$  &    $+44~07~14$  &  $3\farcs13 \times 1\farcs73$  &15.1  & 376.61  &  $1\farcs25 \times 1\farcs10$  & 7.9  & 31.23\\
           NGC\,4254   &   $12~18~49.4$  &    $+14~24~59$  &  $1\farcs99 \times 1\farcs81$  &15.3  & 577.12  &  $1\farcs53 \times 1\farcs11$  & 9.0  & 28.58\\
           NGC\,4321   &   $12~22~54.9$  &    $+15~49~21$  &  $1\farcs89 \times 1\farcs72$  &15.0  & 622.65  &  $1\farcs53 \times 1\farcs17$  & 9.8  & 29.55\\
           NGC\,4536   &   $12~34~27.1$  &    $+02~11~17$  &  $2\farcs16 \times 1\farcs86$  &15.5  & 524.37  &  $2\farcs44 \times 1\farcs45$  &10.2  & 15.58\\
           NGC\,4559   &   $12~35~57.7$  &    $+27~57~36$  &  $1\farcs75 \times 1\farcs63$  &13.6  & 646.04  &  $1\farcs32 \times 1\farcs27$  & 9.0  & 29.13\\
           NGC\,4569   &   $12~36~49.8$  &    $+13~09~46$  &  $2\farcs01 \times 1\farcs80$  &15.4  & 572.68  &  $1\farcs68 \times 1\farcs18$  &10.7  & 29.15\\
           NGC\,4579   &   $12~37~43.6$  &    $+11~49~02$  &  $2\farcs12 \times 1\farcs85$  &23.4  & 807.73  &  $1\farcs72 \times 1\farcs21$  &50.9  &132.87\\
           NGC\,4594   &   $12~39~59.4$  &    $-11~37~23$  &  $2\farcs76 \times 1\farcs74$  &19.7  & 554.75  &  $2\farcs01 \times 1\farcs09$  &13.4  & 32.88\\
           NGC\,4625   &   $12~41~52.4$  &    $+41~16~24$  &  $1\farcs73 \times 1\farcs56$  &13.5  & 677.51  &  $1\farcs54 \times 1\farcs17$  & 9.7  & 29.19\\
           NGC\,4631   &   $12~42~05.9$  &    $+32~32~22$  &  $1\farcs85 \times 1\farcs76$  &14.3  & 597.52  &  $2\farcs29 \times 1\farcs83$  &12.9  & 16.71\\
           NGC\,4725   &   $12~50~26.6$  &    $+25~30~06$  &  $1\farcs85 \times 1\farcs76$  &13.5  & 561.64  &  $1\farcs33 \times 1\farcs26$  & 9.0  & 29.20\\
           NGC\,4736   &   $12~50~53.0$  &    $+41~07~14$  &  $1\farcs72 \times 1\farcs55$  &13.9  & 706.09  &  $2\farcs25 \times 2\farcs06$  &13.4  & 15.67\\
           NGC\,4826   &   $12~56~43.9$  &    $+21~41~00$  &  $1\farcs97 \times 1\farcs75$  &14.3  & 559.38  &  $1\farcs46 \times 1\farcs33$  &10.1  & 28.20\\
           NGC\,5055   &   $13~15~49.2$  &    $+42~01~49$  &  $1\farcs76 \times 1\farcs60$  &12.8  & 619.07  &  $1\farcs51 \times 1\farcs17$  & 9.7  & 29.58\\
           NGC\,5194   &   $13~29~52.7$  &    $+47~11~43$  &  $1\farcs79 \times 1\farcs57$  &17.0  & 820.75  &  $1\farcs54 \times 1\farcs15$  & 9.3  & 28.43\\
           NGC\,5398   &   $14~01~20.2$  &    $-33~04~09$  &  $6\farcs62 \times 1\farcs76$  &20.6  & 239.46  &  $3\farcs98 \times 1\farcs03$  & 8.4  & 11.09\\
           NGC\,5457   &   $14~03~12.6$  &    $+54~20~57$  &  $1\farcs82 \times 1\farcs67$  &13.9  & 619.05  &  $1\farcs55 \times 1\farcs10$  & 8.6  & 27.29\\
           NGC\,5474   &   $14~05~01.3$  &    $+53~39~44$  &  $1\farcs83 \times 1\farcs71$  &14.3  & 616.05  &  $1\farcs49 \times 1\farcs07$  &11.9  & 40.40\\
           NGC\,5713   &   $14~40~11.3$  &    $-00~17~27$  &  $2\farcs55 \times 1\farcs77$  &17.6  & 526.76  &  $2\farcs35 \times 1\farcs80$  &15.4  & 19.67\\
           NGC\,5866   &   $15~06~29.5$  &    $+55~45~48$  &  $1\farcs92 \times 1\farcs63$  &13.8  & 595.03  &  $1\farcs62 \times 1\farcs20$  &15.4  & 42.92\\
           NGC\,6946   &   $20~34~52.3$  &    $+60~09~14$  &  $2\farcs03 \times 1\farcs64$  &16.1  & 655.75  &  $2\farcs15 \times 1\farcs12$  & 9.5  & 21.44\\
           NGC\,7331   &   $22~37~04.1$  &    $+34~24~56$  &  $1\farcs82 \times 1\farcs67$  &15.1  & 672.97  &  $1\farcs51 \times 1\farcs21$  &10.0  & 29.80\\
           NGC\,7793   &   $23~57~49.2$  &    $-32~35~24$  &  $4\farcs77 \times 1\farcs62$  &14.0  & 245.06  &  $4\farcs47 \times 1\farcs08$  & 9.4  & 10.50  
\enddata
\tablecomments{See \citet{ejm18a} for the 33\,GHz imaging characteristics.}
\end{deluxetable*}

\begin{deluxetable*}{l|cc|ccc|ccc}
\tablecaption{Extranuclear Source Positions and Imaging Characteristics \label{tbl-3}}
\tabletypesize{\scriptsize}
\tablewidth{0pt}
\tablehead{
\colhead{} & \colhead{} & \colhead{} & \multicolumn{3}{|c|}{3\,GHz} & \multicolumn{3}{c}{15\,GHz} \\
\colhead{Enuc.\,ID} & \colhead{R.A.} & \colhead{Decl.} &  \colhead{Synthesized}& \colhead{$\sigma$}& \colhead{$\sigma_{T_{\rm b}}$} & \colhead{Synthesized}& \colhead{$\sigma$}& \colhead{$\sigma_{T_{\rm b}}$}\\
\colhead{}  & \colhead{(J2000)} & \colhead{(J2000)} & \colhead{Beam}& \colhead{($\mu$Jy\,bm$^{-1}$)}& \colhead{(mK)} & \colhead{Beam}& \colhead{($\mu$Jy\,bm$^{-1}$)}& \colhead{(mK)}
}
\startdata
  NGC\,0628~Enuc.\,1   &   $01~36~45.1$  &    $+15~47~51$  &  $1\farcs96 \times 1\farcs78$  &14.0  & 542.77  &  $1\farcs56 \times 1\farcs21$  & 9.8  &28.16\\
  NGC\,0628~Enuc.\,2   &   $01~36~37.5$  &    $+15~45~12$  &  $1\farcs96 \times 1\farcs78$  &14.0  & 544.62  &  $1\farcs57 \times 1\farcs22$  & 9.6  &27.27\\
  NGC\,0628~Enuc.\,3   &   $01~36~38.8$  &    $+15~44~25$  &  $1\farcs96 \times 1\farcs78$  &14.0  & 542.12  &  $1\farcs57 \times 1\farcs23$  & 9.6  &26.82\\
  NGC\,0628~Enuc.\,4   &   $01~36~35.5$  &    $+15~50~11$  &  $1\farcs96 \times 1\farcs78$  &13.8  & 537.98  &  $1\farcs56 \times 1\farcs20$  & 8.0  &23.07\\
  NGC\,1097~Enuc.\,1   &   $02~46~23.9$  &    $-30~17~51$  &  $5\farcs76 \times 1\farcs80$  &41.2  & 537.94  &  $3\farcs76 \times 0\farcs95$  &11.8  &17.84\\
  NGC\,1097~Enuc.\,2   &   $02~46~14.4$  &    $-30~15~05$  &  $5\farcs76 \times 1\farcs80$  &38.3  & 500.07  &  $3\farcs80 \times 0\farcs98$  &11.4  &16.42\\
  NGC\,2403~Enuc.\,1   &   $07~36~45.5$  &    $+65~37~00$  &  $2\farcs21 \times 1\farcs53$  &13.7  & 548.00  &  $1\farcs86 \times 1\farcs16$  & 9.4  &23.61\\
  NGC\,2403~Enuc.\,2   &   $07~36~52.7$  &    $+65~36~46$  &  $2\farcs21 \times 1\farcs53$  &13.7  & 547.71  &  $1\farcs85 \times 1\farcs15$  & 9.5  &23.96\\
  NGC\,2403~Enuc.\,3   &   $07~37~06.9$  &    $+65~36~39$  &  $2\farcs21 \times 1\farcs53$  &13.7  & 548.41  &  $1\farcs84 \times 1\farcs16$  & 9.7  &24.51\\
  NGC\,2403~Enuc.\,4   &   $07~37~17.9$  &    $+65~33~46$  &  $2\farcs21 \times 1\farcs53$  &13.7  & 546.14  &  $1\farcs80 \times 1\farcs15$  & 9.5  &24.86\\
  NGC\,2403~Enuc.\,5   &   $07~36~19.5$  &    $+65~37~04$  &  $2\farcs21 \times 1\farcs53$  &13.5  & 541.72  &  $1\farcs79 \times 1\farcs15$  & 9.5  &24.84\\
  NGC\,2403~Enuc.\,6   &   $07~36~28.5$  &    $+65~33~50$  &  $2\farcs21 \times 1\farcs53$  &13.5  & 541.23  &  $1\farcs77 \times 1\farcs13$  & 9.7  &26.15\\
  NGC\,2976~Enuc.\,1   &   $09~47~07.8$  &    $+67~55~52$  &  $2\farcs83 \times 1\farcs66$  &14.5  & 415.03  &  $1\farcs99 \times 1\farcs13$  & 9.8  &23.62\\
  NGC\,2976~Enuc.\,2   &   $09~47~24.1$  &    $+67~53~57$  &  $2\farcs83 \times 1\farcs66$  &14.4  & 412.17  &  $1\farcs97 \times 1\farcs13$  &10.0  &24.19\\
  NGC\,3521~Enuc.\,1   &   $11~05~46.3$  &    $-00~04~10$  &  $3\farcs07 \times 1\farcs95$  &19.4  & 436.89  &  $1\farcs92 \times 1\farcs22$  &14.0  &32.41\\
  NGC\,3521~Enuc.\,2   &   $11~05~49.9$  &    $-00~03~40$  &  $3\farcs07 \times 1\farcs95$  &19.8  & 445.88  &  $2\farcs05 \times 1\farcs11$  &11.7  &27.66\\
  NGC\,3521~Enuc.\,3   &   $11~05~47.6$  &    $+00~00~33$  &  $3\farcs07 \times 1\farcs95$  &18.9  & 425.62  &  $2\farcs21 \times 1\farcs13$  &11.6  &25.02\\
  NGC\,3627~Enuc.\,1   &   $11~20~16.2$  &    $+12~57~50$  &  $1\farcs95 \times 1\farcs77$  &15.4  & 604.26  &  $1\farcs50 \times 1\farcs22$  & 9.0  &26.70\\
  NGC\,3627~Enuc.\,2   &   $11~20~16.3$  &    $+12~58~44$  &  $1\farcs95 \times 1\farcs77$  &15.7  & 614.25  &  $1\farcs92 \times 1\farcs41$  &15.7  &31.48\\
  NGC\,3627~Enuc.\,3   &   $11~20~16.0$  &    $+12~59~52$  &  $1\farcs95 \times 1\farcs77$  &15.7  & 616.15  &  $1\farcs51 \times 1\farcs17$  &13.4  &40.92\\
  NGC\,3938~Enuc.\,1   &   $11~52~46.4$  &    $+44~07~01$  &  $3\farcs13 \times 1\farcs73$  &15.0  & 374.91  &  $1\farcs27 \times 1\farcs12$  & 8.3  &31.38\\
  NGC\,3938~Enuc.\,2   &   $11~53~00.0$  &    $+44~07~55$  &  $3\farcs13 \times 1\farcs73$  &14.8  & 368.03  &  $1\farcs27 \times 1\farcs12$  & 8.3  &31.40\\
  NGC\,4254~Enuc.\,1   &   $12~18~49.1$  &    $+14~23~59$  &  $1\farcs99 \times 1\farcs81$  &15.1  & 567.30  &  $1\farcs55 \times 1\farcs14$  & 8.5  &26.05\\
  NGC\,4254~Enuc.\,2   &   $12~18~44.6$  &    $+14~24~25$  &  $1\farcs99 \times 1\farcs81$  &15.3  & 575.22  &  $1\farcs51 \times 1\farcs16$  & 9.6  &29.55\\
  NGC\,4321~Enuc.\,1   &   $12~22~58.9$  &    $+15~49~35$  &  $1\farcs89 \times 1\farcs72$  &14.9  & 620.02  &  $1\farcs53 \times 1\farcs18$  & 9.8  &29.30\\
  NGC\,4321~Enuc.\,2   &   $12~22~49.8$  &    $+15~50~29$  &  $1\farcs89 \times 1\farcs72$  &14.9  & 618.78  &  $1\farcs52 \times 1\farcs18$  & 9.9  &29.98\\
  NGC\,4631~Enuc.\,1   &   $12~41~40.8$  &    $+32~31~51$  &  $1\farcs87 \times 1\farcs75$  &14.0  & 577.02  &  $1\farcs66 \times 1\farcs27$  & 9.1  &23.44\\
  NGC\,4631~Enuc.\,2   &   $12~42~21.3$  &    $+32~33~07$  &  $1\farcs85 \times 1\farcs76$  &13.9  & 578.32  &  $1\farcs61 \times 1\farcs28$  & 9.0  &23.48\\
  NGC\,4736~Enuc.\,1   &   $12~50~56.2$  &    $+41~07~20$  &  $1\farcs72 \times 1\farcs55$  &13.9  & 705.86  &  $1\farcs54 \times 1\farcs17$  &12.9  &38.89\\
  NGC\,5055~Enuc.\,1   &   $13~15~58.0$  &    $+42~00~26$  &  $1\farcs76 \times 1\farcs60$  &13.1  & 630.58  &  $1\farcs50 \times 1\farcs17$  & 9.9  &30.50\\
  NGC\,5194~Enuc.\,1   &   $13~29~53.1$  &    $+47~12~40$  &  $1\farcs79 \times 1\farcs57$  &16.8  & 810.49  &  $1\farcs53 \times 1\farcs14$  & 9.1  &28.20\\
  NGC\,5194~Enuc.\,2   &   $13~29~44.1$  &    $+47~10~21$  &  $1\farcs74 \times 1\farcs52$  &17.0  & 863.15  &  $1\farcs52 \times 1\farcs14$  & 8.9  &27.62\\
  NGC\,5194~Enuc.\,3   &   $13~29~44.6$  &    $+47~09~55$  &  $1\farcs74 \times 1\farcs52$  &16.9  & 858.25  &  $1\farcs52 \times 1\farcs14$  & 9.1  &28.37\\
  NGC\,5194~Enuc.\,4   &   $13~29~56.2$  &    $+47~14~07$  &  $1\farcs79 \times 1\farcs57$  &16.0  & 768.85  &  $1\farcs52 \times 1\farcs14$  & 9.5  &29.61\\
  NGC\,5194~Enuc.\,5   &   $13~29~59.6$  &    $+47~14~01$  &  $1\farcs79 \times 1\farcs57$  &16.1  & 773.77  &  $1\farcs52 \times 1\farcs14$  & 9.7  &30.21\\
  NGC\,5194~Enuc.\,6   &   $13~29~39.5$  &    $+47~08~35$  &  $1\farcs74 \times 1\farcs52$  &15.8  & 805.61  &  $1\farcs51 \times 1\farcs12$  & 8.7  &27.81\\
  NGC\,5194~Enuc.\,7   &   $13~30~02.5$  &    $+47~09~52$  &  $1\farcs74 \times 1\farcs52$  &17.8  & 904.22  &  $1\farcs52 \times 1\farcs11$  & 9.3  &29.80\\
  NGC\,5194~Enuc.\,8   &   $13~30~01.6$  &    $+47~12~52$  &  $1\farcs79 \times 1\farcs57$  &16.7  & 803.48  &  $1\farcs58 \times 1\farcs18$  & 9.7  &28.04\\
  NGC\,5194~Enuc.\,9   &   $13~29~59.9$  &    $+47~11~12$  &  $1\farcs79 \times 1\farcs57$  &17.0  & 819.52  &  $1\farcs59 \times 1\farcs16$  &10.7  &31.20\\
 NGC\,5194~Enuc.\,10   &   $13~29~56.7$  &    $+47~10~46$  &  $1\farcs74 \times 1\farcs52$  &18.1  & 921.23  &  $1\farcs59 \times 1\farcs16$  &10.8  &31.55\\
 NGC\,5194~Enuc.\,11   &   $13~29~49.7$  &    $+47~13~29$  &  $1\farcs79 \times 1\farcs57$  &16.3  & 783.23  &  $1\farcs57 \times 1\farcs21$  &12.8  &36.27\\
  NGC\,5457~Enuc.\,1   &   $14~03~10.2$  &    $+54~20~58$  &  $1\farcs82 \times 1\farcs67$  &13.8  & 615.51  &  $1\farcs55 \times 1\farcs10$  & 8.4  &26.65\\
  NGC\,5457~Enuc.\,2   &   $14~02~55.0$  &    $+54~22~27$  &  $1\farcs81 \times 1\farcs63$  &14.1  & 643.39  &  $1\farcs54 \times 1\farcs09$  & 8.5  &27.14\\
  NGC\,5457~Enuc.\,3   &   $14~03~41.3$  &    $+54~19~05$  &  $1\farcs82 \times 1\farcs67$  &14.1  & 627.94  &  $4\farcs68 \times 2\farcs20$  &25.0  &13.13\\
  NGC\,5457~Enuc.\,4   &   $14~03~53.1$  &    $+54~22~06$  &  $1\farcs82 \times 1\farcs73$  &13.8  & 590.56  &  $1\farcs54 \times 1\farcs08$  & 9.5  &30.71\\
  NGC\,5457~Enuc.\,5   &   $14~03~01.1$  &    $+54~14~29$  &  $1\farcs81 \times 1\farcs63$  &13.8  & 629.72  &  $1\farcs49 \times 1\farcs07$  &10.8  &36.48\\
  NGC\,5457~Enuc.\,6   &   $14~02~28.1$  &    $+54~16~26$  &  $1\farcs86 \times 1\farcs67$  &13.7  & 596.38  &  $1\farcs53 \times 1\farcs06$  &11.9  &39.68\\
  NGC\,5457~Enuc.\,7   &   $14~04~29.3$  &    $+54~23~46$  &  $1\farcs82 \times 1\farcs73$  &13.7  & 589.02  &  $1\farcs50 \times 1\farcs04$  &13.0  &44.82\\
  NGC\,5713~Enuc.\,1   &   $14~40~12.1$  &    $-00~17~47$  &  $2\farcs55 \times 1\farcs77$  &17.6  & 525.60  &  $2\farcs35 \times 1\farcs80$  &15.9  &20.36\\
  NGC\,5713~Enuc.\,2   &   $14~40~10.5$  &    $-00~17~47$  &  $2\farcs55 \times 1\farcs77$  &17.6  & 527.29  &  $2\farcs35 \times 1\farcs80$  &15.9  &20.36\\
  NGC\,6946~Enuc.\,1   &   $20~35~16.6$  &    $+60~10~57$  &  $2\farcs03 \times 1\farcs64$  &15.2  & 616.55  &  $2\farcs07 \times 1\farcs10$  & 8.3  &19.75\\
  NGC\,6946~Enuc.\,2   &   $20~35~25.1$  &    $+60~10~03$  &  $2\farcs03 \times 1\farcs64$  &15.0  & 609.01  &  $2\farcs00 \times 1\farcs09$  & 9.8  &24.32\\
  NGC\,6946~Enuc.\,3   &   $20~34~52.2$  &    $+60~12~41$  &  $2\farcs03 \times 1\farcs64$  &15.0  & 608.66  &  $2\farcs09 \times 1\farcs09$  & 8.7  &20.65\\
  NGC\,6946~Enuc.\,4   &   $20~34~19.4$  &    $+60~10~09$  &  $2\farcs03 \times 1\farcs64$  &14.8  & 603.01  &  $2\farcs45 \times 1\farcs90$  &10.1  &11.78\\
  NGC\,6946~Enuc.\,5   &   $20~34~39.0$  &    $+60~04~53$  &  $2\farcs06 \times 1\farcs63$  &15.1  & 604.38  &  $1\farcs91 \times 1\farcs11$  & 8.8  &22.34\\
  NGC\,6946~Enuc.\,6   &   $20~35~06.0$  &    $+60~11~01$  &  $2\farcs03 \times 1\farcs64$  &15.7  & 636.59  &  $1\farcs95 \times 1\farcs11$  & 8.6  &21.28\\
  NGC\,6946~Enuc.\,7   &   $20~35~11.2$  &    $+60~08~60$  &  $2\farcs03 \times 1\farcs64$  &15.7  & 636.59  &  $2\farcs14 \times 1\farcs13$  & 8.8  &19.64\\
  NGC\,6946~Enuc.\,8   &   $20~34~32.2$  &    $+60~10~20$  &  $2\farcs05 \times 1\farcs63$  &15.8  & 639.98  &  $1\farcs99 \times 1\farcs11$  & 8.7  &21.20\\
  NGC\,6946~Enuc.\,9   &   $20~35~12.7$  &    $+60~08~53$  &  $2\farcs03 \times 1\farcs64$  &15.7  & 636.59  &  $2\farcs14 \times 1\farcs13$  & 8.8  &19.64\\
  NGC\,7793~Enuc.\,1   &   $23~57~48.8$  &    $-32~36~59$  &  $4\farcs77 \times 1\farcs62$  &13.9  & 243.82  &  $4\farcs62 \times 1\farcs09$  & 9.6  &10.26\\
  NGC\,7793~Enuc.\,2   &   $23~57~56.1$  &    $-32~35~40$  &  $4\farcs77 \times 1\farcs62$  &14.0  & 244.33  &  $4\farcs71 \times 1\farcs11$  & 9.8  &10.09\\
  NGC\,7793~Enuc.\,3   &   $23~57~48.8$  &    $-32~34~53$  &  $4\farcs77 \times 1\farcs62$  &14.0  & 245.29  &  $4\farcs47 \times 1\farcs08$  & 8.8  & 9.84  
\enddata
\tablecomments{See \citet{ejm18a} for the 33\,GHz imaging characteristics.}
\end{deluxetable*}

\vspace{-0.3cm}

\subsection{Interferometric Imaging}

Calibrated VLA measurement sets for each source were imaged using the task {\sc tclean} in CASA version 4.6.0. For some cases, the Ka-band images contain data from observations taken during both the 11B and 13A semesters, but are heavily weighted by the 13A semester observations, as those include significantly more data. The mode of {\sc tclean} was set to multi-frequency synthesis \citep[{\sc mfs};][]{mfs1,mfs2}. We chose a pixel scale of $0\farcs2$ for all three bands, and adopted Briggs weighting with {\sc robust }$=0.5$ and {\sc nterms }$=2$. This allows the cleaning procedure to also model the spectral index variations on the sky. Although this procedure utilizes the large fractional bandwidths of each observation to generate in-band spectral index maps, we do not use them in our analysis given that the signal-to-noise ratio ($S/N$) of our sources is typically too low for them to be reliable. To help deconvolve extended low-intensity emission, we took advantage of the multiscale {\sc clean} option \citep{msclean,msmfs} in CASA, searching for structures with scales $\sim$1 and 3 times the FWHM of the synthesized beam. The choice of our final imaging parameters was the result of extensive experimentation to identify values that yielded the best combination of brightness-temperature sensitivity and reduction of artifacts resulting from strong sidelobes in the naturally weighted beam for these snapshot-like observations.

A primary beam correction was applied using the CASA task {\sc impbcor} before analyzing the images. The primary-beam-corrected continuum images at $3-33$GHz are shown in Figure 1.~The FWHM of the synthesized beam along with the corresponding point-source and brightness temperature sensitivities for each image are given in Tables 2 and 3. Finally, in order to accurately compare the flux density measured for each star-forming region across the full $3-33$\,GHz frequency range, we use the CASA task {\sc imsmooth} to match all VLA images for each pointing to a common circularized, Gaussian, beam corresponding to the lowest angular resolution among all three frequency bands scaled by a factor of $\sqrt{2}\times0\farcs2$ (i.e., the pixel scale). This scaling factor eliminates cases for the convolution kernels used by {\sc imsmooth} to have length zero in any axis in units of pixels. We additionally crop all images to a common field-of-view (FOV) equal to the FOV of our 33\,GHz observations (i.e., a primary beam FWHM of 44\arcsec).

\begin{figure*}
\center
	\includegraphics[scale=0.26]{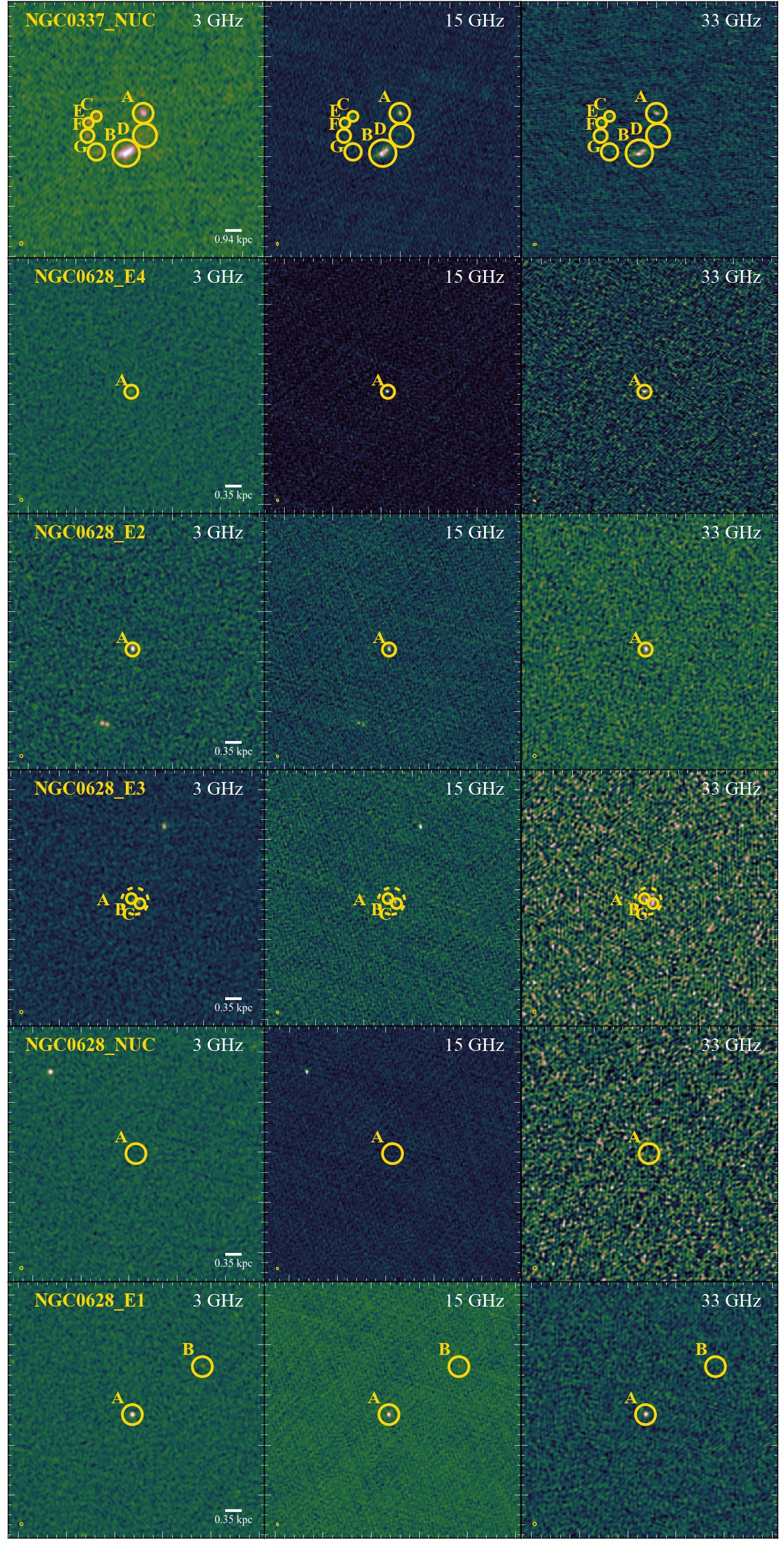}
\caption{3-panel images for all 56 galaxies in the SFRS showing the combination of our 3 GHz (left), 15 GHz (middle) and 33 GHz (right) observations. The color scale \citep{green11} is set to one of 4 power-law stretches: [a(p-p$_{min}$)/(p$_{max}$-p$_{min}$)], where p is the pixel value and $a=1/3$, 0.5, 1.0, and 2.0. A cube-root stretch of $a=1/3$ was used when the brightest pixel in the image had a $S/N>500$. A square-root stretch of $a=0.5$ was used when the brightest pixel in the image had an $50<S/N<500$. A linear stretch was used when the brightest pixel was between $10<S/N<50$, and a square stretch was used when the brightest pixel had a $S/N<10$. A scalebar of $10\arcsec$ is also given in the bottom right corner of each panel. To distinguish between individual sources identified in the full-resolution and 7\arcsec~smoothed maps, we use uppercase and lowercase letters as part of their names for reporting photometry in Tables 4 and 5, respectfully. An extended version of this Figure is available in the Appendix.}
\end{figure*}

\section{Ancillary Data and Photometry}

In this section we provide a description of ancillary data used, as well as our procedure for extracting consistent aperture photometry across the full suite of multi-wavelength data available for galaxies in the SFRS.

\vspace{-2mm}
\subsection{Ancillary Data}

%By including data across the electromagnetic spectrum, we can make use of the different timescales of the emission mechanisms for thermal, nonthermal, UV, and IR flux to estimate the star-formation history (SFH) within individual star-forming regions in nearby galaxies. Rest-frame UV (125-250 nm) and optical (e.g., H$\alpha$) observations provide a measure of the energetic photons released from the ionized gas surrounding newly formed massive stars ($t \leq 10$ Myr). Whereas, rest-frame IR measurements, particularly those in the mid- and far-infrared, do not suffer significantly from extinction, and can accurately probe the total SFR (including the contribution from evolved low-mass stellar populations in star-forming regions). Further, since free-free emission is only produced by the shortest-lived massive stars, its presence in large quantities relative to synchrotron emission is indicative of very young star-formation. These timescales suggest that regions with flat spectral indices, and enhanced IR-low-frequency radio flux density ratios ($q_{IR}$) also host the youngest areas of ongoing star-formation. These correlations can then be used to study the evolution of star-formation activity across a large sample of star-forming galaxies which show large variations in their local ISM conditions. An updated calibration of the coefficients found for the hybrid (H$\alpha$ + 24$\mu$m and UV + 24$\mu$m) SFR equations presented in Calzetti et al. (2007) and subsequently updated in Hao et al. (2011), will be presented in a forthcoming paper (Dong et al. 2020 in prep). 

GALEX far-UV (FUV; 1528\AA) and near-UV (NUV; 2271\AA) data were taken from the GALEX Large Galaxy Atlas \citep{sei07}. The calibration uncertainty for these data is $\sim 15\%$ in both bands. One galaxy, NGC\,1377, does not have existing near- or far-UV imaging.

The H$\alpha$ data used in this analysis is taken from the compilation presented in \citet{akl12}, where details about the data quality and preparation (e.g., correction for [NII] emission) can be found, as well as the SINGS archive. All H$\alpha$ images were then further corrected for foreground stars. The typical resolution of the seeing-limited H$\alpha$ images is $\sim 1-2\arcsec$, and the calibration uncertainty among these maps is taken to be $\sim 20\%$. Two galaxies, IC\,342 and NGC\,2146 do not have H$\alpha$ imaging from SINGS.
%, and are therefore excluded from all plots and correlations presented below which rely on the H$\alpha$ imaging.

Archival \textit{Spitzer} 8$\mu$m and 24$\mu$m data were largely taken from the SINGS and Local Volume Legacy (LVL) legacy programs, and have a calibration uncertainty of $\sim 5\%$. Details on the associated observation strategies and data reduction steps can be found in \citet{dd07, dd09}. Two galaxies, IC\,342 and NGC\,2146, were not a part of SINGS or LVL; their 24 $\mu$m imaging comes from \citet{ce08}.

Finally, in order to account for the significant differences between the point-spread functions (PSF) of the various telescopes used in this analysis, we implement the convolution kernels presented in \citet{ga11} to convolve the instrumental PSF for each image, and produce a corrected Gaussian beam of 7\arcsec~at each wavelength.

\subsection{Region Identification and Aperture Photometry}

To identify and characterize potential star-forming regions in the SFRS, we start by searching within an area that is equal to twice the FWHM of the VLA primary beam at 33\,GHz ($\sim 160$"). Any region visually identified in least one radio band is retained for further investigation. The location of all radio-identified sources were compared with the 8$\,\mu$m maps of each galaxy to determine if the source has a corresponding detection in the mid-IR, and is thus likely associated with a star-forming region. Sources characterized as potential background galaxies have no obvious 8\,$\mu$m counterpart, and very rarely a 33\,GHz counterpart. In total, we visually identified 389 regions for which we perform aperture photometry to determine those that are statistically significant detections.~Of the 389 sources visually identified, 377 had a statistically significant detection (i.e., $S/N > 3$) in at least one band, which is given in Table 4. For completeness, the remaining 12 sources are still labeled in the panels of Figure 1 to demonstrate the full identification process we utilized.

% as well as the ground-based H$\alpha$ line flux, which have the same approximate resolution and PSF shape. These measurements allow us to determine the thermal fractions and H$\alpha$-to-33\,GHz flux ratios for individual H{\sc ii} regions identified in the SFRS.
Using the CASA task {\sc imstat}, we measured and report the $3-33$\,GHz flux densities for each region. The size of the apertures were hand-selected to fully encompass the visible extent of the $3-33$\,GHz radio flux density of each region, with an additional constraint of having a diameter equal to or larger than the FWHM of the synthesized beam for that pointing. 
%takes into account differences in the synthesized beam position angles among the different images.  
We do not apply aperture corrections to our photometry given that we both convolve all images for a given pointing to common beam and use the same sized aperture at all wavelengths. Photometric uncertainties were conservatively estimated by taking the empirically measured noise from empty regions in each non primary beam corrected image, applying the empirical primary beam correction based on Equation 4 in EVLA Memo 195, and scaling by the ratio of the synthesized beam area to the adopted aperture area. This noise is then added in quadrature with the VLA calibration uncertainty \citep[{$\sim$3\%};][]{pb13fcal}. The median size of the apertures used for all 377 sources is $164 \pm 6.3$pc with a median absolute deviation of 97\,pc. The aperture sizes used for each identified region is given in Table 4. 
%\textbf{We do not include aperture corrections to in our photometry given that we convolve all data to a common Gaussian beam and use the same apertures for all bands.  Furthermore, the thermal fraction and spectral index measurements rely on relative differences between the flux density measurements.}

We additionally carried out photometry on the full VLA, GALEX, H$\alpha$, and \textit{Spitzer}/MIPS 24\,$\mu$m datasets after matching their resolution (7\arcsec~FWHM), cropping each image to a common field of view, and re-gridding to a common pixel scale ($0\farcs2$).~In total, we identify 180 discrete sources, and critical apertures of 7\arcsec~were used in all cases. Unlike our native-resolution photometry, we report sources with upper-limits in all radio bands if a statistically significant detection exists in our ancillary GALEX and \textit{Spitzer} imaging.~The median size of the apertures used at 7\arcsec~resolution is $259 \pm 7.2$\,pc with a median absolute deviation of 73\,pc.~The UV and H$\alpha$ photometry of each region was corrected for Milky Way extinction using \citet{ds98} assuming $A_{V}/E(B - V) = 3.1$ and the modeled extinction curves of \citet{wd01} and \citet{bd03}.~The results from our radio photometry are given in Table 5.~The GALEX, H$\alpha$, and \textit{Spitzer}/MIPS 24\,$\mu$m photometry results, which are presented in Table 6, are not used directly in the present analysis, but will be utilized for further studies. 

Finally, in order to account for and remove any diffuse emission component that is most likely unassociated with the most recent star formation activity in the disks of these galaxies, we measure a local background value within the vicinity of each star-forming region.~The local background was measured by placing an annulus a distance of 1.5 times the synthesized beam FWHM away from the center of the source position in both the full-resolution and smoothed maps. The median surface brightness within this annulus was then multiplied by the effective area of the beam to get an estimate of the local diffuse background emission. These values are given in Table 7. Further, we measure the fractional contribution of the background emission for each region, and find that the median value is $4.7 \pm 0.43 \%$, $6.6 \pm 0.79 \%$, and $3.9 \pm 0.71 \%$, with median absolute deviations of $5.3\%$, $5.4\%$, and $4.5\%$ for our 3, 15, and 33\,GHz observations respectively. Importantly, these values are smaller than the $15-40\%$ found for the regions studied at $25\arcsec$ ($\sim 1$kpc) scales with the GBT \citep{ejm11b}. 

The regions, listed in Tables 4 and 5, are named according to the nearest 33\,GHz image, with an alphabetical suffix if there are multiple regions corresponding to one image. For example, ``NGC\,2403\,Enuc.\,2\,B"  is the second of three regions within the VLA pointing of extranuclear region 2 in NGC\,2403. We distinguish individual sources identified in the 7\arcsec~smoothed maps by instead using a lowercase letter. For example, ``NGC\,2403 Enuc.\,2\,b" is one of two regions in the image of extranuclear region 2 in NGC\,2403, and is composed of the sum contribution of ``NGC\,2403\,Enuc.\,2\,B" and ``NGC\,2403\,Enuc.\,2\,C" in the full-resolution maps.

\startlongtable
\begin{deluxetable*}{lcc|ccc|cc|cc}
\tablecolumns{10}
\tablecaption{Region Photometry and Derived Parameters \label{tbl-5}}
\tablewidth{0pt}
\tablehead{
\colhead{Source ID} & \colhead{R.A.} & \colhead{Decl.} & \colhead{$S_\mathrm{3\,GHz}$} & \colhead{$S_\mathrm{15\,GHz}$} & \colhead{$S_\mathrm{33\,GHz}$ }&   \colhead{$d_{\rm ap}$} & \colhead{$r_\mathrm{G}$} & \colhead{$\alpha$}  & \colhead{$f^{\rm 33\,GHz}_{\rm T}$}\\
\colhead{} & \colhead{(J2000)} & \colhead{(J2000)} & \colhead{(mJy)}  & \colhead{(mJy)}  & \colhead{(mJy)} & \colhead{(pc)} & \colhead{(kpc)} &  \colhead{}  & \colhead{} 
}
\startdata
\cutinhead{Star-Forming Regions}
                            NGC\,0337\,D   &             00~59~49.94  &            -07~34~47.60  & $  1.19\pm  0.07$  & $  0.77\pm  0.04$  & $  0.31\pm  0.05$  &  655.  & 0.921  & $ -0.335\pm  0.042$  & $  0.841\pm  0.038$  \\
                            NGC\,0337\,A   &             00~59~50.01  &            -07~34~34.33  & $  0.99\pm  0.06$  & $  0.70\pm  0.03$  & $  0.28\pm  0.05$  &  561.  & 0.881  & $ -0.278\pm  0.042$  & $  0.888\pm  0.033$  \\
                            NGC\,0337\,B   &             00~59~50.69  &            -07~34~58.26  & $  4.94\pm  0.07$  & $  2.66\pm  0.04$  & $  1.80\pm  0.07$  &  749.  & 2.041  & $ -0.399\pm  0.012$  & $  0.780\pm  0.012$  \\
                            NGC\,0337\,C   &             00~59~51.89  &            -07~34~36.20  & $  0.30\pm  0.03$  & $  0.29\pm  0.02$  & $  0.21\pm  0.03$  &  281.  & 3.735  & $ -0.076\pm  0.060$  & $  1.000\pm  0.028$  \\
                            NGC\,0337\,G   &             00~59~51.89  &            -07~34~57.50  & $  0.36\pm  0.05$  & $  0.19\pm  0.03$  & $  0.17\pm  0.05$  &  468.  & 3.123  & $ -0.358\pm  0.102$  & $  0.820\pm  0.096$  \\
                            NGC\,0337\,E   &             00~59~52.22  &            -07~34~40.10  & $  0.28\pm  0.03$  & $  0.27\pm  0.02$  & $  0.30\pm  0.03$  &  281.  & 4.048  & $  0.022\pm  0.059$  & $  1.000\pm  0.022$  \\
                            NGC\,0337\,F   &             00~59~52.26  &            -07~34~47.90  & $  0.28\pm  0.04$  & $  0.19\pm  0.02$  & $  0.22\pm  0.04$  &  374.  & 3.753  & $ -0.159\pm  0.087$  & $  0.968\pm  0.051$  \\
                     NGC\,0628\,Enuc.\,4   &             01~36~35.69  &            +15~50~07.22  & $  0.27\pm  0.03$  & $  0.26\pm  0.02$  & $  0.23\pm  0.03$  &  140.  & 7.614  & $ -0.043\pm  0.065$  & $  1.000\pm  0.029$  \\
\enddata
\tablecomments{$^{\dagger}$\,The aperture used for this region contained multiple smaller individual regions.}
\end{deluxetable*}

\clearpage

\startlongtable
\begin{deluxetable*}{lcc|ccc|cc|cc}
\tablecolumns{10}
\tablecaption{Region Photometry and Derived Parameters at 7\arcsec~Angular Resolution \label{tbl-6}}
\tablewidth{0pt}
\tablehead{
\colhead{Source ID} & \colhead{R.A.} & \colhead{Decl.} & \colhead{$S_\mathrm{3\,GHz}$} & \colhead{$S_\mathrm{15\,GHz}$} & \colhead{$S_\mathrm{33\,GHz}$ }&   \colhead{$d_{\rm ap}$} & \colhead{$r_\mathrm{G}$} & \colhead{$\alpha$}  & \colhead{$f^{\rm 33\,GHz}_{\rm T}$}\\
\colhead{} & \colhead{(J2000)} & \colhead{(J2000)} & \colhead{(mJy)}  & \colhead{(mJy)}  & \colhead{(mJy)} & \colhead{(pc)} & \colhead{(kpc)} &  \colhead{}  & \colhead{} 
}
\startdata
\cutinhead{Star-Forming Regions}
                            NGC\,0337\,a   &            00~59~50.018  &             -07~34~33.9  & $  0.53\pm  0.02$  & $  0.37\pm  0.01$  & $  0.23\pm  0.01$  &  655.  & 0.945  & $ -0.30\pm  0.03$  & $  0.87\pm  0.02$  \\
                            NGC\,0337\,b   &            00~59~50.683  &             -07~34~57.6  & $  1.70\pm  0.05$  & $  0.91\pm  0.03$  & $  0.61\pm  0.02$  &  655.  & 1.966  & $ -0.42\pm  0.02$  & $  0.76\pm  0.02$  \\
                            NGC\,0337\,c   &            00~59~51.995  &             -07~34~54.9  & $  0.13\pm  0.01$  & $  0.12\pm  0.01$  & $  0.10\pm  0.02$  &  655.  & 3.227  & $ -0.10\pm  0.07$  & $  1.00\pm  0.03$  \\
                            NGC\,0337\,d   &            00~59~52.156  &             -07~34~38.2  & $  0.25\pm  0.02$  & $  0.25\pm  0.01$  & $  0.21\pm  0.02$  &  655.  & 4.061  & $ -0.05\pm  0.04$  & $  1.00\pm  0.02$  \\
                     NGC\,0628\,Enuc.\,4   &            01~36~35.718  &            +15~50~07.25  & $  0.14\pm  0.01$  & $  0.13\pm  0.01$  & $  0.12\pm  0.01$  &  244.  & 7.608  & $ -0.04\pm  0.04$  & $  1.00\pm  0.02$  \\
                     NGC\,0628\,Enuc.\,2   &            01~36~37.645  &             +15~45~07.2  & $  0.26\pm  0.01$  & $  0.15\pm  0.01$  & $  0.18\pm  0.01$  &  244.  & 4.468  & $ -0.22\pm  0.03$  & $  0.93\pm  0.02$  \\
                     NGC\,0628\,Enuc.\,3   &            01~36~38.779  &             +15~44~23.2  & $  0.27\pm  0.01$  & $  0.21\pm  0.01$  & $  0.19\pm  0.02$  &  244.  & 5.720  & $ -0.16\pm  0.03$  & $  0.97\pm  0.02$  \\
                               NGC\,0628   &              01~36~41.7  &               +15~46~59  &         $<$  0.03  &         $<$  0.02  &         $<$  0.04  &  244.  & 0.071  &           \nodata  &           \nodata  \\
\enddata
\end{deluxetable*}

\section{Results}

%IMPORTANT: The number of regions is 332 SF and 377 Total. 12 are not detected in any radio bands, and therefore don't meet the stated requirement, but are in the lists.

Using the 3, 15, and 33\,GHz photometry, along with the 8\,$\mu$m imaging from \textit{Spitzer}, we classify each region as either a star-forming region (SF), a background galaxy candidate (BG), a likely supernova remnant (SNe/R: see Section 5.4), or an anomalous microwave emission candidate (AME: see Section 5.5). In total we have identified 320 star-forming regions, 14 likely background galaxies, 10 likely supernovae/supernova remnants, and 33 AME candidates. Given that we are primarily interested in emission arising from our sample galaxies, the potential background galaxies have been removed from all plots, and are discussed as a separate population of sources in Section 5.3. Regions identified at 7\arcsec~which include emission from potential AME and SNe/R candidates are correspondingly classified in Table 5.

We present results for the spectral index and thermal fraction distributions only including regions identified in the SFRS that have a $S/N \geq 3$ measured in at least two radio bands. This ensures that our fit results are not biased by single detections at one frequency, and allows us to make accurate comparisons with the regions identified in M18a, for which a detection at 33\,GHz was required. This requirement removes 4 likely background galaxies, 4 likely supernova remnants, and 34 star-forming regions. These single-band detections are distributed almost uniformly across all three frequency bands: 19, 11, and 12 at 3, 15, and 33\,GHz, respectively. Accordingly, in the sample used to study the radio spectral indices and thermal fractions, we have retained 335 (286 SF, 10 BG, 6 SNe/R, and 33 AME) of the 377 sources with a statistically significant detection in at least one band. The median $S/N$ of these 335 sources is 18, 15, and 9 for detections at 3, 15, and 33\,GHz, respectively.

%0 likely background galaxies, 0 likely supernova remnants, and
%in the sample used to study the radio spectral indices and thermal fractions with the 7\arcsec~smoothed images,
We apply the same criteria for sources included in the spectral index and thermal fraction analysis using the 7\arcsec~smoothed images. This removes 17 star-forming regions. These single-band detections are also distributed uniformly across all three bands: 7, 4, and 6 at 3, 15, and 33\,GHz, respectively. 
Accordingly, we have retained 163 (142 SF, 0 BG, 1 SNe/R, and 20 AME) of the 180 sources with a statistically significant detection in at least one band. The median $S/N$ of these 163 sources is 24, 23, and 19 for detections at 3, 15, and 33\,GHz, respectively.

\subsection{Spectral Indices}

The simplest approach to modeling the radio spectra of galaxies is by adopting a two-component power-law, with the thermal/nonthermal ratio as well as the non-thermal spectral index allowed to vary as free parameters. For many star-forming galaxies in the local Universe, this model adequately describes the dominant physical processes occurring at radio frequencies \citep{jc92}. However, a robust interpretation of the radio spectrum can be complex. For example: the thermal and nonthermal fractions may vary with galaxy mass \citep[e.g.][]{ah07,efb03}, the nonthermal index can vary within galaxies \citep{fat17}, and AME may add an additional component to the radio spectra at high frequencies in some regions \citep[e.g.][]{ejm10,ejm18b}.

\begin{figure}
\centering
  \includegraphics[scale=0.4]{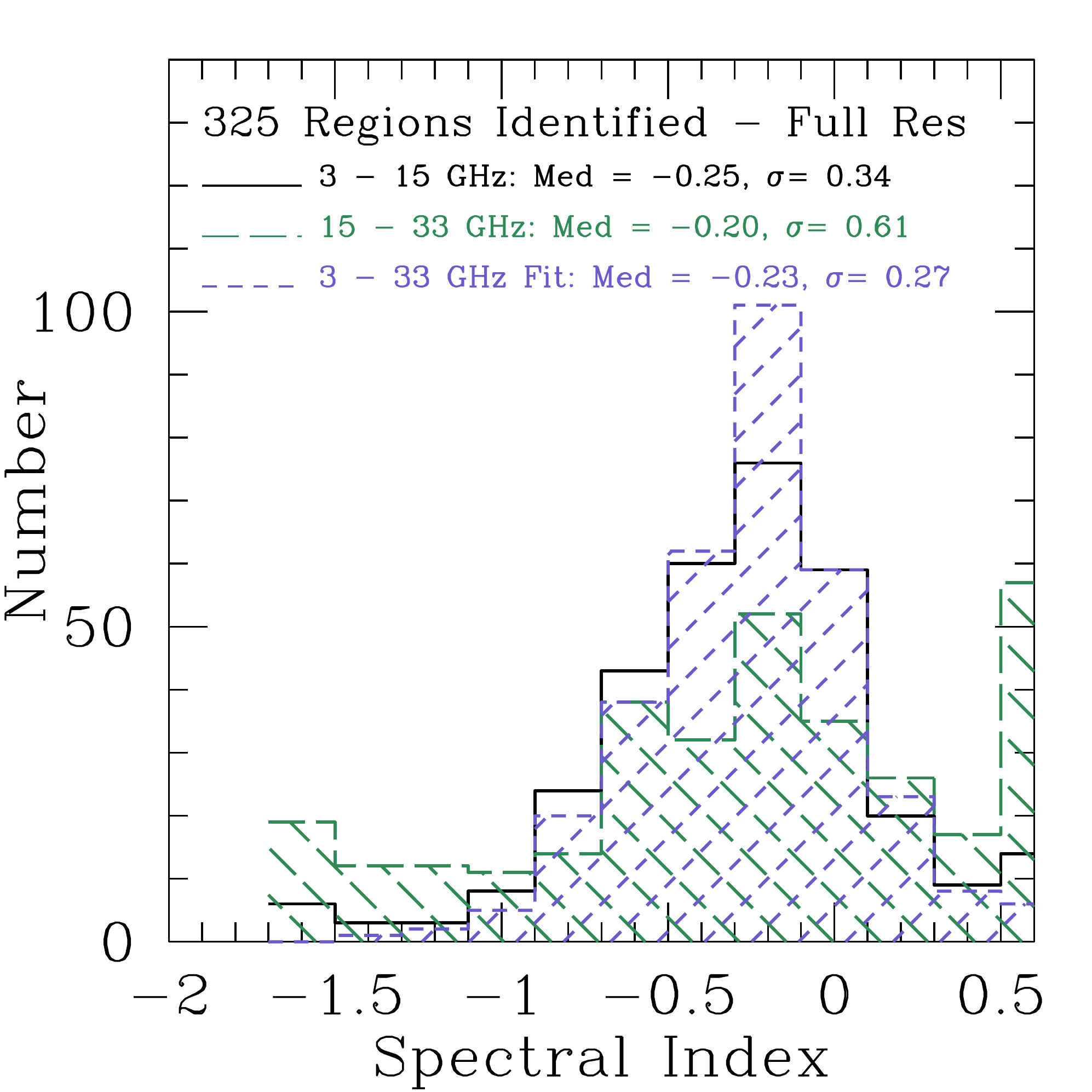}
  \includegraphics[scale=0.4]{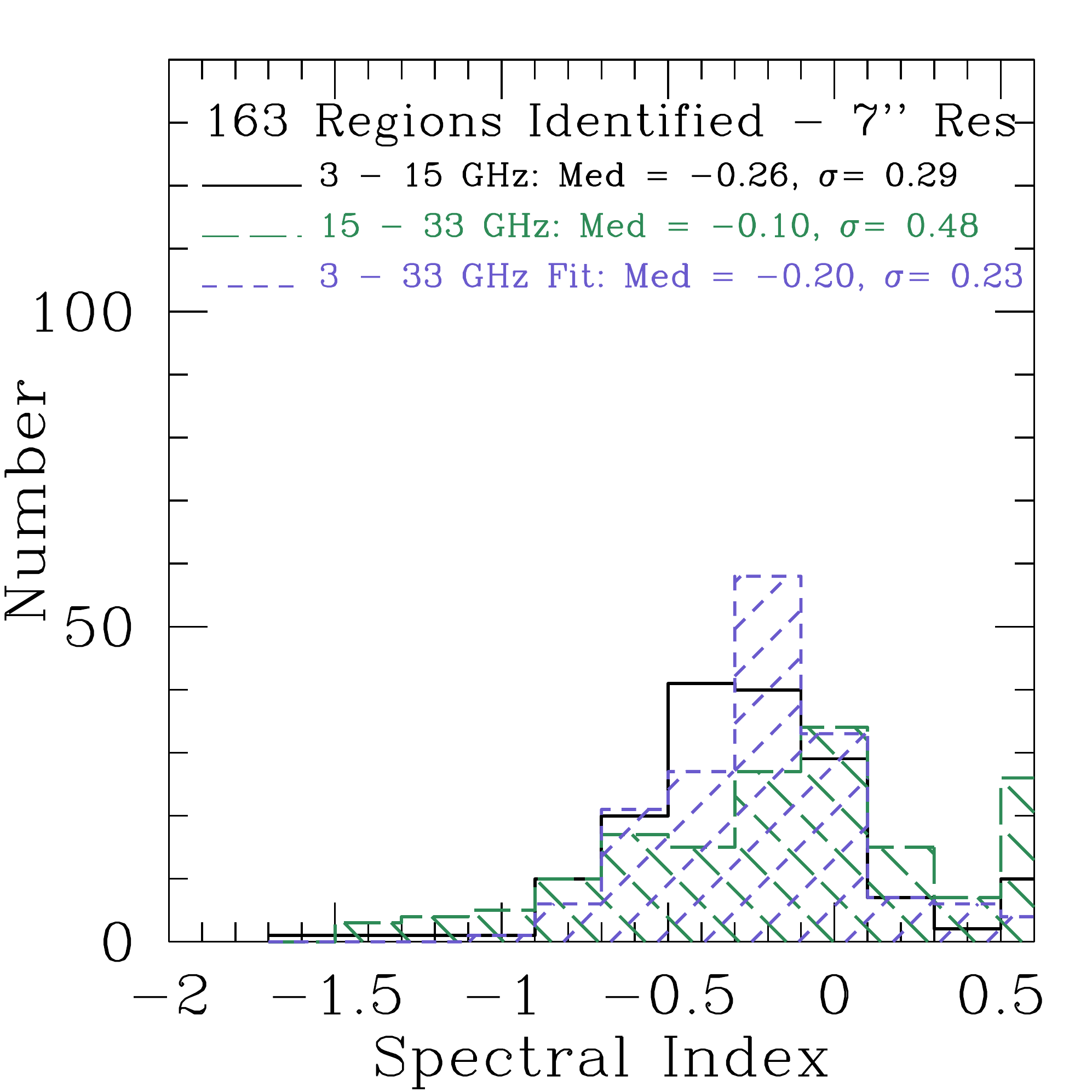}
\caption{Top: The $3-15$\,GHz (black), $15-33$\,GHz (green), and $3-33$\,GHz (purple) radio spectral index distributions for 325 regions identified in the SFRS. The median size of the apertures used is $162 \pm 6.5$pc. Bottom: The spectral index distributions for 163 7\arcsec~regions identified in M18a.~The median size of the apertures used at 7\arcsec~resolution is $259 \pm 7.2$\,pc Overall, we find that the $3-33$\,GHz distributions measured at $\sim2\arcsec$ and 7\arcsec~~are consistent with one another, implying that free-free emission dominates the radio spectra of star-forming regions on scales up to $\sim 500$pc. Not included in any of these plots are the likely background galaxy candidates identified.}
\end{figure}

To measure the $3 - 33$\,GHz spectral indices, we performed a linear least-squares fit to the data with a single power-law representing the combination of thermal and nonthermal emission. The distributions of the measured spectral slopes with the 10 likely background galaxies removed are given in the top panel of Figure 2 (325 total regions). The median spectral indices we measure from $3 -15$, $15 - 33$, and $3 - 33$\,GHz are $-0.25 \pm 0.024$, $-0.20 \pm 0.043$, and $-0.23 \pm 0.018$, respectively. The median absolute deviation of these distributions is 0.34, 0.61, and 0.27 respectively. Interestingly, we do not see particularly steep spectral indices from $3-15$\,GHz, indicating that the contribution of non-thermal emission to the radio flux density of individual star-forming regions is marginal on $\sim 100$ pc scales in these galaxies. This is consistent with the results presented in \citet{ejm12b}, where they show that the thermal fraction at 33\,GHz increased as function of decreasing linear resolution. Despite the relatively flat spectral slopes from $3-15$\,GHz, we do see evidence that the spectrum continues to flatten from $15-33$\,GHz, on average. This is consistent with expectations for star-forming regions, where the radio spectrum is synchrotron-dominated at low frequencies, and flattens at higher frequencies as the contribution of thermal emission increases \citep{jc12,msc10,ejm13}. Finally, we do not see any significant evidence for free-free absorption, which is known to affect the compact central regions of local star-forming and starburst galaxies, and would result in steep spectral indices even at high frequencies \citep{cy90,msc08,ejm13}.

In the nearby Universe, the typical radio spectrum of normal star-forming galaxies is well-described by a power-law spectrum with a spectral index of $-0.7$, and a thermal fraction of $\sim10\%$ at $\sim$ 1\,GHz \citep{kwm88,cy90}, whereas studies of local luminous infrared galaxies (LIRGs) find a flat spectrum around 1\,GHz and a steepening spectrum above 10\,GHz \citep{msc08, akl11b,ejm13}. New high-resolution observations of a large sample of LIRGs in the Great Observatories All-Sky LIRG Survey (GOALS) have revealed that the steep spectrum seen in highly star-forming galaxies are attributed solely to the nucleus, and that in extranuclear regions the spectral shape is typical of the star-forming regions identified in this study \citep{stl19}.

%%Add sentences about CR diffusion, or maybe an entire section?
%This interpretation can be complicated by the fact that the low-frequency radio emission is tracing cosmic rays, which potentially diffuse out of the photometric apertures used as they lose energy. However, Murphy et al. (2006) showed that for star-forming regions with measured star formation rate surface densities of $-1.5 <$log$(\Sigma_{SFR})< -1$ (i.e., the distribution of extra-nuclear star-forming regions studied here: See the following Section), the mean free path length of a cosmic ray is $\sim 100$pc (corresponding to a $\tau_{cool} \sim 1$x$10^{5}$); this is still five times smaller than the physical scale of our aperture in the closest galaxy in the sample (NGC0838). Thus we do not expect CR diffusion to cause suppression of the observed synchrotron emission.

\subsection{Thermal Fractions}
Given the results above, we now calculate the thermal fraction of each region by using the spectral index, measured in Section 4.1, from $3-15$\,GHz ($\alpha_{3-15\,{\rm GHz}}$) to set the lower-limit on the nonthermal spectral index ($\alpha^{\rm NT}$) such that  $\alpha^{\rm NT} = -0.83$ if $\alpha_{3-15\,{\rm GHz}} \geq -0.83$, and $\alpha^{\rm NT} = \alpha_{3-15\,{\rm GHz}}$ if $\alpha_{3-15\,GHz} < -0.83$. A constant nonthermal radio spectral index of $-0.83$ is assumed based on the average non-thermal spectral index found among the 10 star-forming regions studied in NGC\,6946 by \citet{ejm11b}. Furthermore, this value is consistent with the results of \citet[][i.e., $\alpha^{\rm NT} = -0.83$ with a scatter of $\sigma_{\alpha^{\rm NT}} = 0.13$]{nb97} for a sample of 74 nearby galaxies. Finally, we adopt a single power-law exponent for the free-free emission ($\sim -0.1$), and use the fit from 3 to 33\,GHz to set the overall radio spectral index. Then, using the prescription in \citet{kwb84}, we can calculate the thermal fraction at 33\,GHz such that, 

\begin{equation} 
f_{\rm T}^{\nu_{1}} = \frac{(\frac{\nu_{2}}{\nu_{1}})^{-\alpha} - (\frac{\nu_{2}}{\nu_{1}})^{-\alpha^{\rm NT}}}{(\frac{\nu_{2}}{\nu_{1}})^{-0.1} - (\frac{\nu_{2}}{\nu_{1}})^{-\alpha^{\rm NT}}}
\end{equation}

\noindent
where $\nu_{1}$ is the target frequency (33 GHz), $\alpha$ is the observed slope from 3 to 33 GHz, and $\alpha^{\rm NT}$ is the nonthermal spectral index. In the top Panel of Figure 3 we show the resulting thermal fractions of the star-forming regions in our sample using the empirically measured values from Equation 1. We find that the median value is $92 \pm 0.8\%$ with a median absolute deviation of $11\%$. This demonstrates that we can reliably use the 33\,GHz flux density to infer the total free-free emission, and thus current star formation activity, on the scales of individual H{\sc ii} and star-forming regions. While this result had been suggested by our previous GBT and VLA campaigns \citep[e.g.,][]{ejm11b,ejm12b,ejm18a}, this is the first measurement of the 33\,GHz thermal fraction based on the shape of the radio spectrum at these frequencies and spatial scales in nearby galaxies.

Further, by restricting our analysis such that we remove all non-SF regions, all apertures containing multiple smaller individual regions, and require $r_{\rm G} \geq 250$pc (238 regions), we mitigate contamination from any central AGN and can determine broad population statistics for extragalactic H{\sc ii} regions using our cleanest sample of extranuclear star-forming regions. In Figure 4 we find that the median thermal fraction is $93 \pm 0.8\%$ with a median absolute deviation of $10\%$. This value is consistent with the results presented in Figures 2, and 3, and confirms that supernova remnants, ultra-compact H{\sc ii} regions, and/or AME candidates do not bias our results.

Finally, we measure the thermal fraction at 33\,GHz for 163 regions identified at 7\arcsec~resolution in M18a to be $94 \pm 0.8$\% with a median absolute deviation of $8\%$. This result is consistent with the measurements at 2" and confirms that free-free emission dominates the radio spectra of star-forming regions on scales up to $\sim$500pc. Further, when we apply the same radial cut (i.e., $r_{\rm G} > 250$pc) and remove all non-SF regions, retaining 111/163 regions, the median thermal fraction at 33\,GHz is $97 \pm 0.5\%$ with a median absolute deviation of $4.6\%$. Thus, it is clear that regardless of the size of the photometric apertures used, isolating extranuclear SF regions in nearby galaxies results in a higher thermal fraction at 33\,GHz and a smaller median absolute deviation in the overall distribution. 

\begin{figure}
\centering
  \includegraphics[scale=0.4]{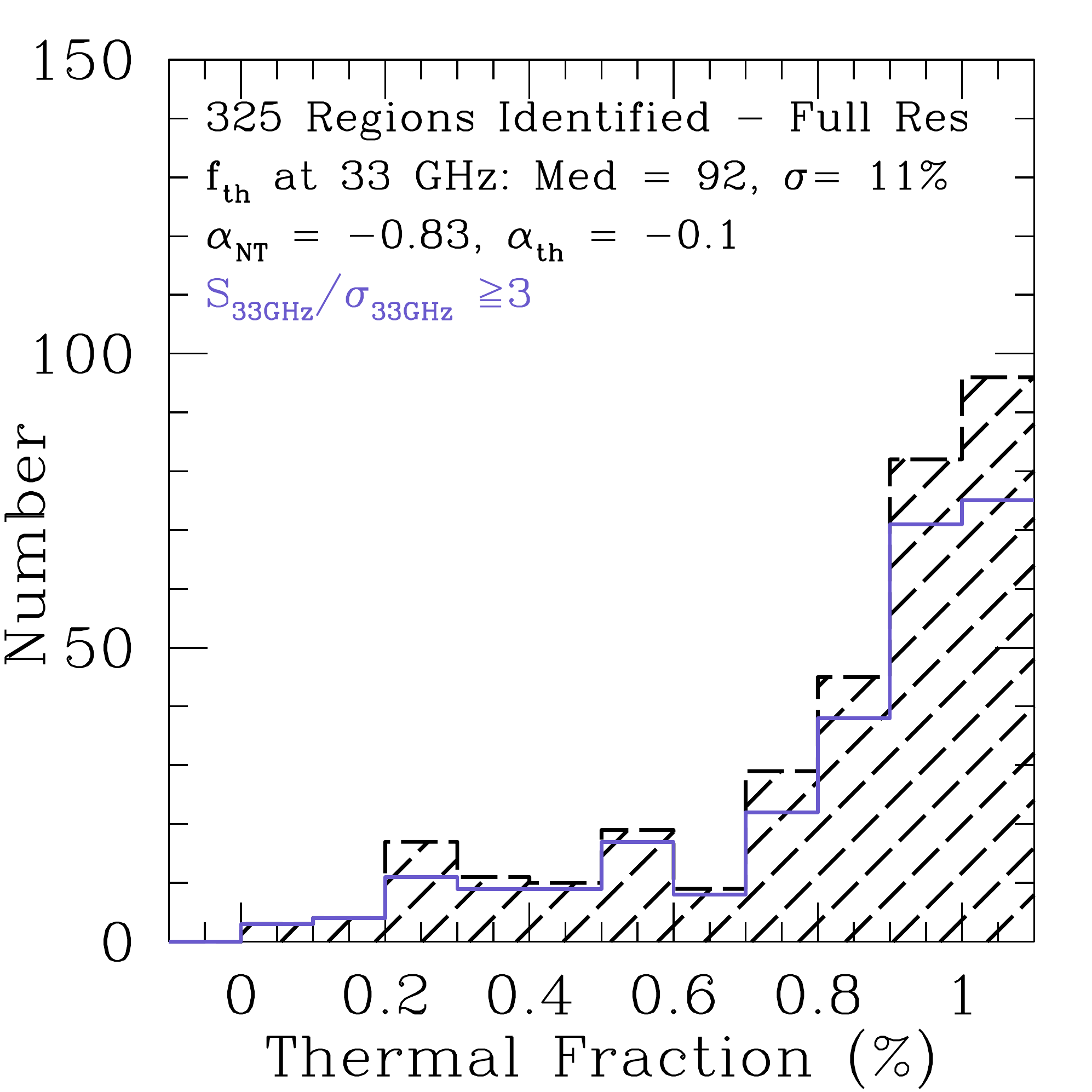}
  \includegraphics[scale=0.4]{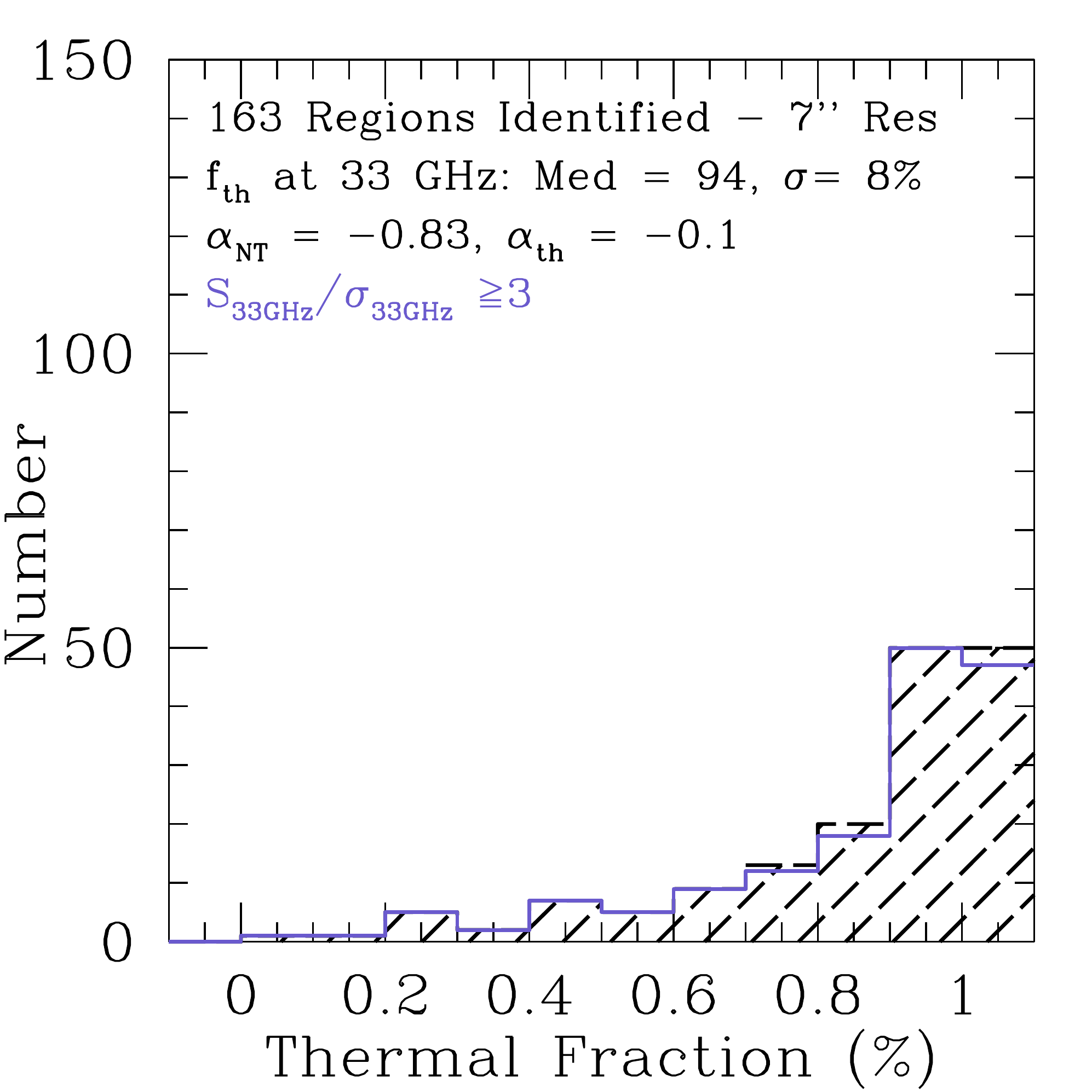}
\caption{Top: The thermal fraction distribution at 33\,GHz for 325 star-forming regions in the SFRS (black). In purple we show the distribution for regions with a $S/N \geq 3$ at 33\,GHz, demonstrating that the lack of a significant 33\,GHz detection does not bias our results. The median size of the apertures used is $162 \pm 6.5$pc. Bottom: The thermal fraction distribution for 163 7\arcsec~regions identified in M18a with a $S/N >$ 3 in two radio bands.~The median size of the apertures used at 7\arcsec~resolution is $259 \pm 7.2$\,pc. Overall, we find that the median thermal fraction at 33\,GHz is $\sim 93\%$, and that this value does not vary significantly from 100 up to $\sim500$pc scales in our galaxy sample.}
\end{figure}

\begin{figure*}[t]
\centering
\resizebox{17cm}{!}{
\includegraphics[scale=1]{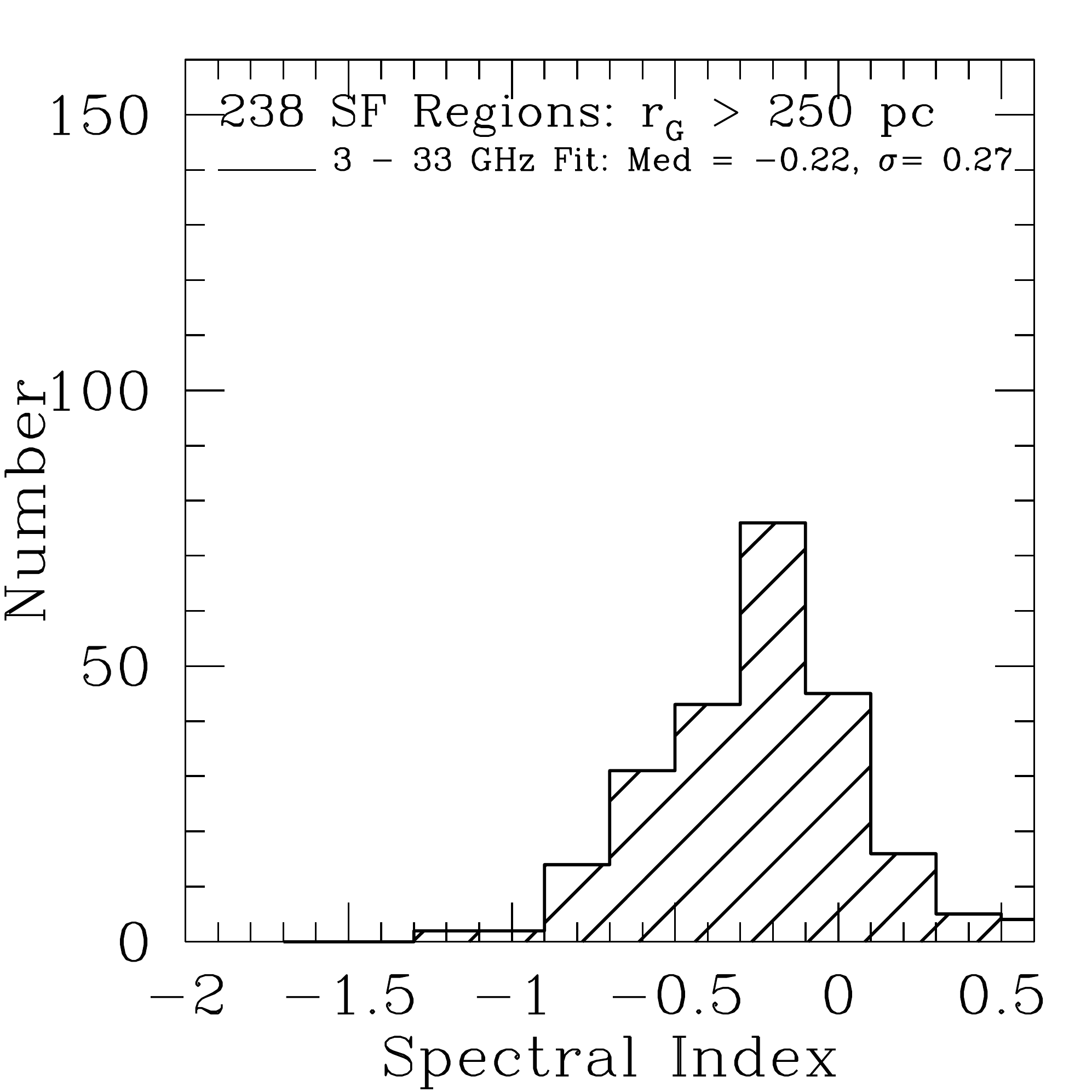}
\includegraphics[scale=1]{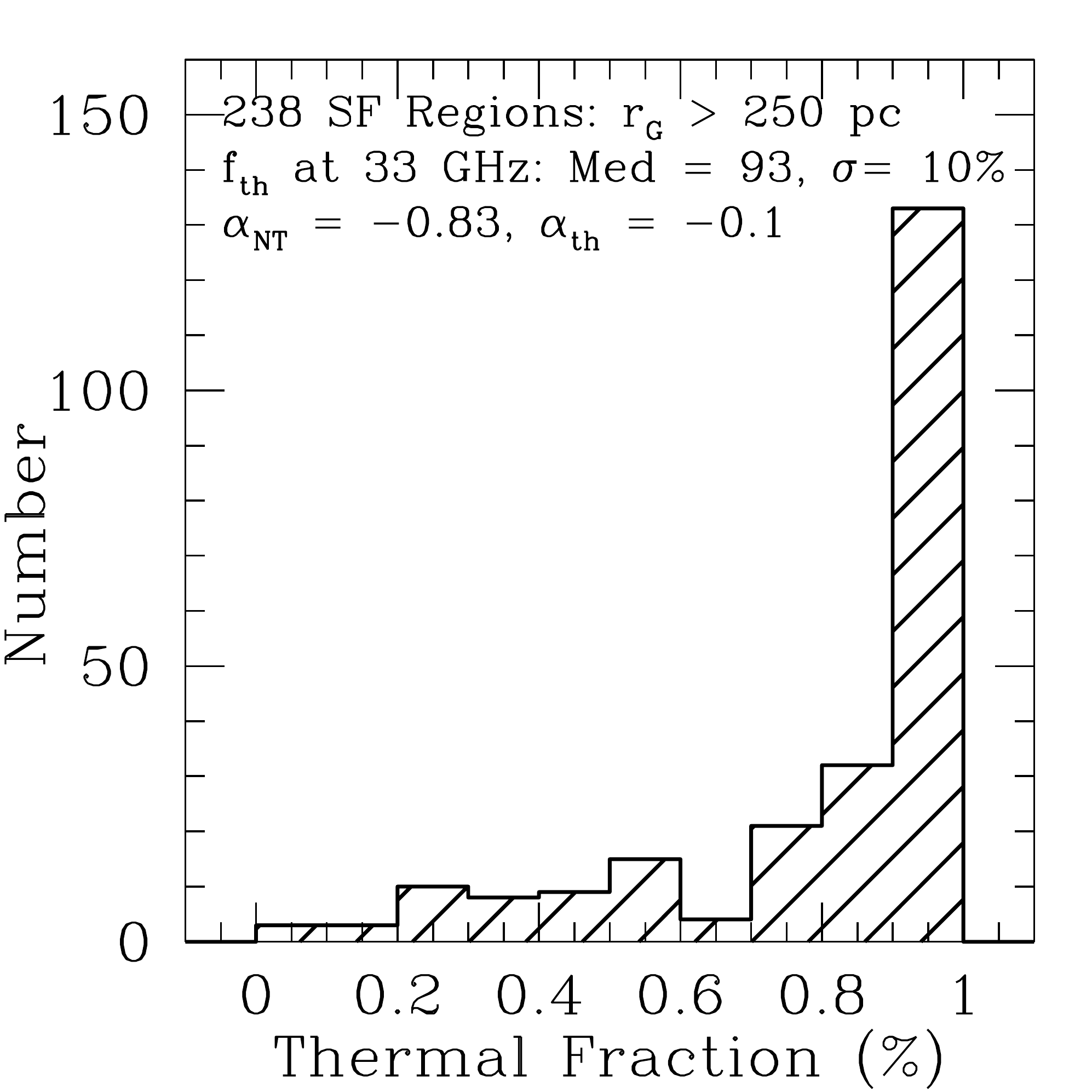}}
\caption{The spectral index and thermal fraction distributions for all 238 SF regions with the likely supernova remnants and AME candidates removed, and a requirement of $r_{\rm G} \geq 250$pc (The same cut adopted in M18a). This represents the cleanest sample of extranuclear star-forming regions for which we can determine broad population statistics for extragalactic H{\sc ii} regions. Ultimately we find that the median spectral index and thermal fraction distributions at 33\,GHz are consistent with the results presented in Figures 2 and 3, confirming that supernova remnants, ultra-compact H{\sc ii} regions, and AME candidates do not bias our results.}
\end{figure*}

\subsection{MCMC Parameter Estimation}
%: Niklas et al. 1997; Callingham et al. 2015; Tabatabaei et al. 2017
%, where the value of the nonthermal spectral index is held fixed,
From Equation 1 it is clear that reliable estimates for the 33\,GHz thermal fraction are sensitive to the value adopted for the nonthermal spectral index ($\alpha^{\rm NT}$). While the median $3-15$\,GHz spectral index observed for our sample is well below the canonical value (i.e., $-0.83$), degeneracies may still exist between $\alpha^{\rm NT}$ and $f_{\rm T}$. In this case classical $\chi^{2}$ fitting methods may underestimate the true uncertainties associated with modeling the radio spectrum as a two-component power-law [i.e., $S(\nu) = A\nu^{\alpha^{\rm NT}}+B\nu^{-0.1}$]. Here, we explore whether the marginalized posterior distributions from a Monte-Carlo Markov Chain (MCMC) analysis better reflect the uncertainties associated with this decomposition \citep{dh10}.

For this exercise, we use the Python package {\sc emcee} \citep{fm13} to generate posterior probability distributions for each of the fitted parameters ($\alpha^{\rm NT}$, $f_{\rm T}$, and $A/B$) given the typical $S/N$ ratio of our three-band observations. Following \citet{jw18}, we parameterize our two-component power-law model at a reference frequency of 1\,GHz to avoid dependencies in frequency space. Further, we adopt a gaussian probability distribution function for the nonthermal spectral index whose mean and standard deviation are consistent with the values obtained in \citet{nb97}. Finally, we make a slight modification to Equation 9 presented in \citet{jw18}, 

\begin{equation} 
P(\theta) \propto H(A, B) e^{\frac{-(\alpha - 0.83)^{2}}{0.13^{2}}}
\end{equation}

\noindent
such that H is equal to 1 when the values of A and B are greater than zero. This is done in order to constrain the nonthermal spectral index, thermal fraction, and normalization constants of each component simultaneously. We make 1000 realizations of this model at three different $S/N$ ratios (5, 10, and 50) by randomly selecting a nonthermal spectral index ($ -2 < \alpha^{\rm NT} < 0$) and vales of A(B) [$ 0 < A (B) < 1$], to represent the typical variance in the values observed for our sample.

In Figure 5 we plot the relative difference in the input and output thermal fraction as a function of the input nonthermal spectral index. For a fixed $S/N$ per region of 10, and with only three data points, our MCMC modeling can recover the input nonthermal spectral index to within 1$\sigma$ for $-1.25 < \alpha^{\rm NT} < 0.25$. Importantly, we find that the best-constrained spectra have nonthermal indices very close to the fixed-value adopted for our $\chi^{2}$-minimization ($-0.83$). This result demonstrates that adopting fixed values for the $\alpha^{\rm NT}$ does not introduce systematic biases into the derived 33\,GHz thermal fractions, over a reasonable set of input conditions for our two-component power-law model. This is an important result for calibrations of the total star-formation rate, which rely on using the observed radio continuum \citep{ejm11b, ejm12b}.

\begin{figure}
\center
	\includegraphics[scale=0.4]{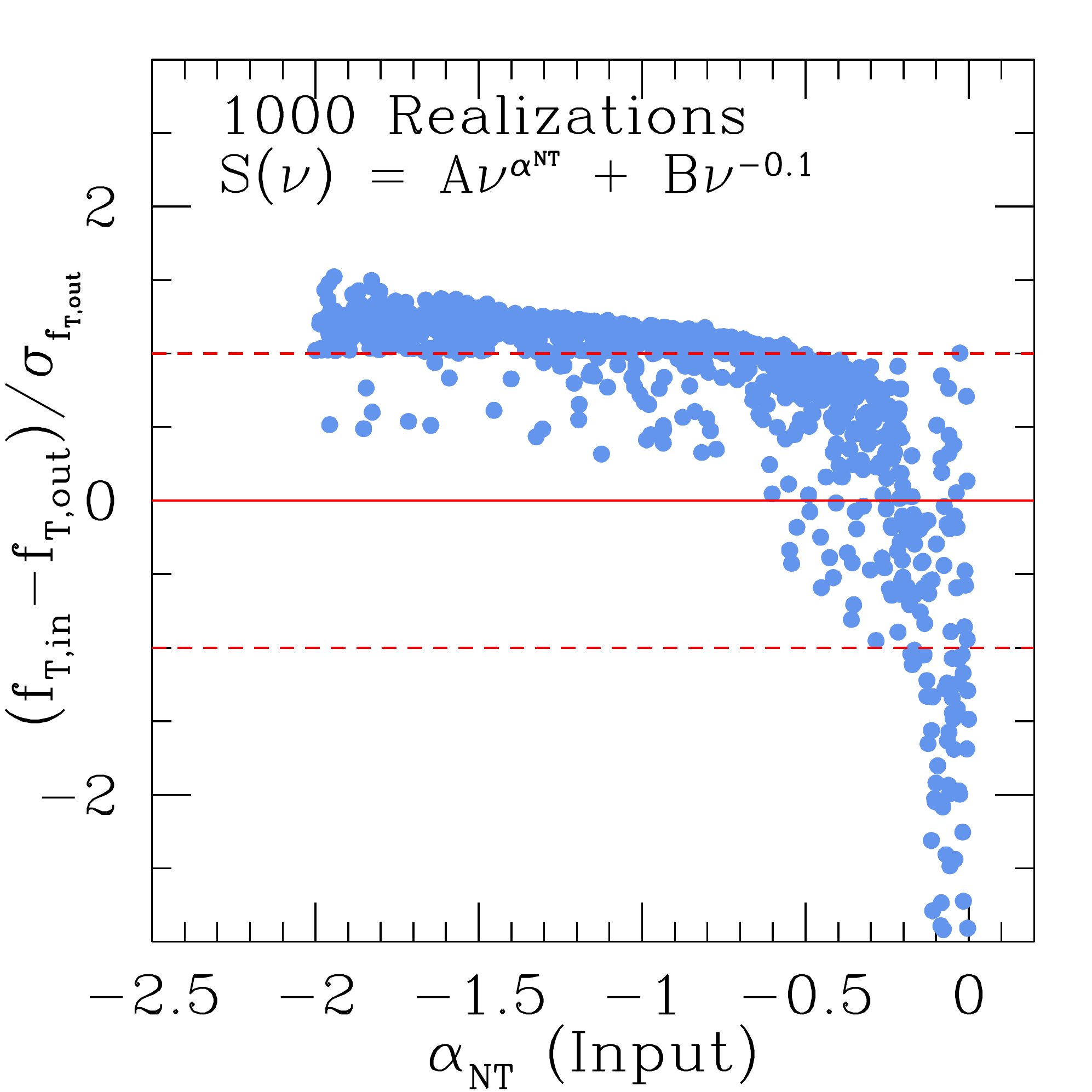}
\caption{The results from our MCMC analysis of 1000 random realizations of 3 - 33\,GHz spectra using as a model $S(\nu) = A\nu^{\alpha^{\rm NT}}+B\nu^{-0.1}$. For these realizations we fix $S/N = 10$ and vary the thermal fraction at 33\,GHz by randomly assigning a non-thermal spectral index from 0 to 2, and an A (B) values between 0 and 1. It is clear from this exercise that by fitting our data using an MCMC approach, our ability to recover the true value for the 33 GHz thermal fraction peaks at $\alpha^{\rm NT,in} \sim -0.8$, which closely resembles the canonical value for the nonthermal spectral index.}
\end{figure}

\section{Discussion}

\subsection{Trends with Galactocentric Radius}

Following the same procedure as M18a, we use the measured position angle (PA) and inclination of each galaxy to convert the angular separation of each star-forming region from the nucleus into a de-projected galactocentric radius ($r_{\rm G}$). In M18a we found that the median 33\,GHz continuum-to-H$\alpha$ line flux ratio was statistically larger within $r_{\rm G} < 250$ pc relative to the outer disk regions by a factor of $1.82 \pm 0.39$, while the ratio of 33\,GHz-to-24$\mu$m flux densities is lower by a factor of $0.45 \pm 0.08$. Such a situation may arise if the circumnuclear regions of these galaxies have extended star formation histories in which star formation that has taken place over a longer period of time, resulting in an accumulation of young dust-heating stars in addition to much older bulge stars that boost the 24\,$\mu$m flux density relative to what is seen in the extranuclear regions. This is largely opposite to what we would expect if there was an additional nonthermal component powering the 33\,GHz emission in the central regions of these galaxies, unless the excess dust-heating at 24\,$\mu$m far exceeds any additional nonthermal emission contribution at 33\,GHz. 

Therefore, these results suggested that the larger ratio of 33\,GHz flux density to H$\alpha$ line flux found in the central regions of these galaxies may primarily arise from increased extinction. We can now test this picture for 325 discrete regions (background galaxies removed) with detailed radio spectral fitting and thermal fraction estimates, which do not suffer from the effects of variable dust extinction in galaxies. In Figure 6 it is clear that the overall dispersion in the measured spectral index and thermal fraction at 33\,GHz increases significantly for regions that lie within the 250\,pc galactocentric radius cut used in M18a to distinguish extranuclear from nuclear/circumnuclear star-forming regions. In fact, limiting the analysis to sources with $r_{\rm G} < 250$\,pc results in a value for the median 33\,GHz thermal fraction of $\sim 71 \pm 3.5\%$ with a median absolute deviation of $\sim $18\% relative to $\sim 95 \pm 2.2\%$ with a median absolute deviation of $\sim $11\% for regions with $r_{\rm G} \geq 250$\,pc. Additionally, the scatter of the thermal fraction distribution increases by nearly a factor of 2 within $r_{\rm G} < 250$pc. This confirms that while extinction may play a role in driving the previously seen correlations, excess nonthermal emission is indeed present in many of the circumnuclear star-forming regions observed in the SFRS.

\begin{figure*}
\center
	\includegraphics[scale=0.6]{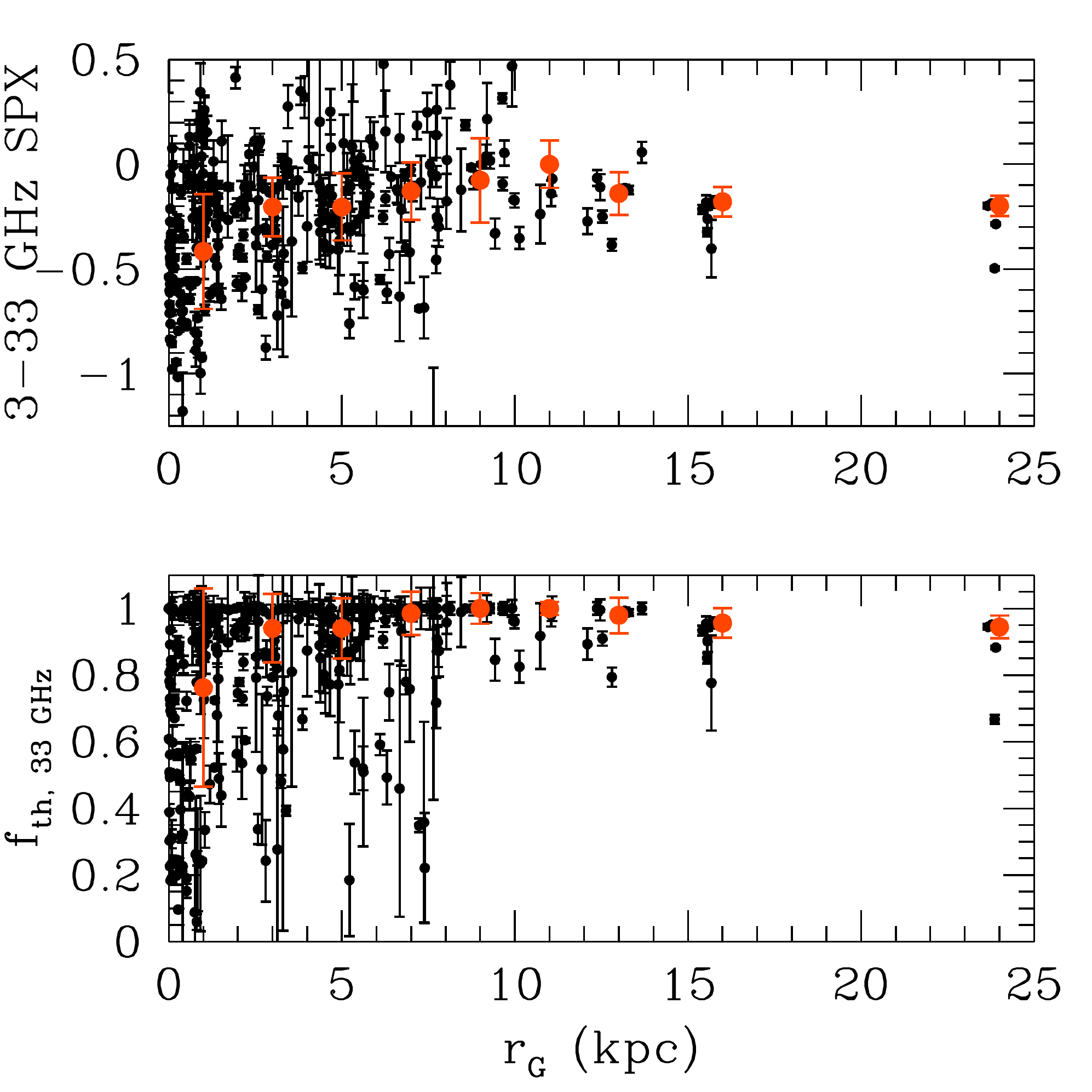}
\caption{The spectral index and 33\,GHz thermal fraction distributions plotted against galactocentric radius for all 325 sources identified in Figure 2. While we identify regions across the full extent of galaxy disks which are heavily dominated by thermal emission, a clear trend emerges where the scatter in both quantities increases significantly as a function of decreasing galactocentric radius (orange points). In particular, no region with a measured $f_{\rm T} < 80\%$ is found in any SFRS galaxy beyond a radius of 7\,kpc. These trends are reflective of the ongoing star-formation activity occurring in the centers of nearby normal galaxies, and reinforce our ability to successfully capture the SFH of individual H{\sc ii} regions using the $3-33$\,GHz radio spectral slopes.}
\end{figure*}

\subsection{Model Age Fitting}

By making use of differences in the timescales associated with thermal (free-free) and synchrotron emission, we can place estimates on the age of star-forming regions by examining how these processes affect the radio spectral indicex from $3 - 33$\,GHz. Since free-free emission is directly associated with ionizing photons that are only produced by the shortest-lived ($\leq 10$ Myr) massive stars, its presence in large quantities relative to synchrotron emission is indicative of very young star formation. 

To better-quantify these different timescales, we use a Starburst 99 (SB99) model of a single instantaneous burst with default inputs (solar metallicity and 2-component Kroupa IMF) run for 1 Gyr \citep{cl99}. However, in order to take into account the fact that at the median physical scales we are probing, ($\sim 100$\,pc), a single, instantaneous starburst may not be representative, we also include SB99 models with a continuous star formation history of SFR = $1\,M_{\odot}$\,yr$^{-1}$ using the same metallicity and IMF input as in the instantaneous burst model (Figure 7 - Left Panel). The details of these models, and their application to extranuclear star-forming regions identified in LIRGs is presented in \citet{stl19}.

For both models we then perform a $\chi^{2}$ minimization to the observed spectral indices for each star-forming region in the sample. We stress that without including cosmic-ray propagation, which will affect the relative ratio of free-free and non-thermal emission on these scales, this exercise is simply meant to understand how the observed distribution of spectral indices can be represented as a distribution of ages. In the left panel of Figure 7 it is clear that our models are insensitive to H{\sc ii} regions with ages $\tau_{H{\sc ii}} < 3$\,Myr and $\tau_{H{\sc ii}} > 40\,$Myr. However, in the intermediate age range we find that a typical uncertainty of $4-6\%$ on the observed spectral index corresponds to a $0.1-0.2$\,dex uncertainty in age. Therefore while the age of any individual region may be uncertain, the increase in the number of young ($3 < \tau_{H{\sc ii}} < 10$\,Myr) relative to old ($\tau_{H{\sc ii}} \sim 20-30$\,Myr) star-forming regions is robust.

The right panel of Figure 7 shows the distribution of fitted ages for the instantaneous burst model in blue and the continuous model in green. Overall, we find that the majority ($\sim 70\%$) of our regions are best-modeled by a continuous burst, with a median age of $\sim 10$\,Myr (grey distribution). Further we find that when all regions are modeled using a continuous SFH, an age-gradient emerges in Figure 6. At small $r_{\rm G}$, the regions with low thermal fractions, which drive the observed scatter, are preferentially older ($\tau_{H{\sc ii}} \sim 20-30$\,Myr). This further supports the notion that the central star-forming regions are, on average, older, and that the associated cosmic ray population in the inner disk of these galaxies is being continuously replenished by ongoing star-formation.

\begin{figure*}[t]
\centering
\resizebox{17cm}{!}{
\includegraphics[scale=1]{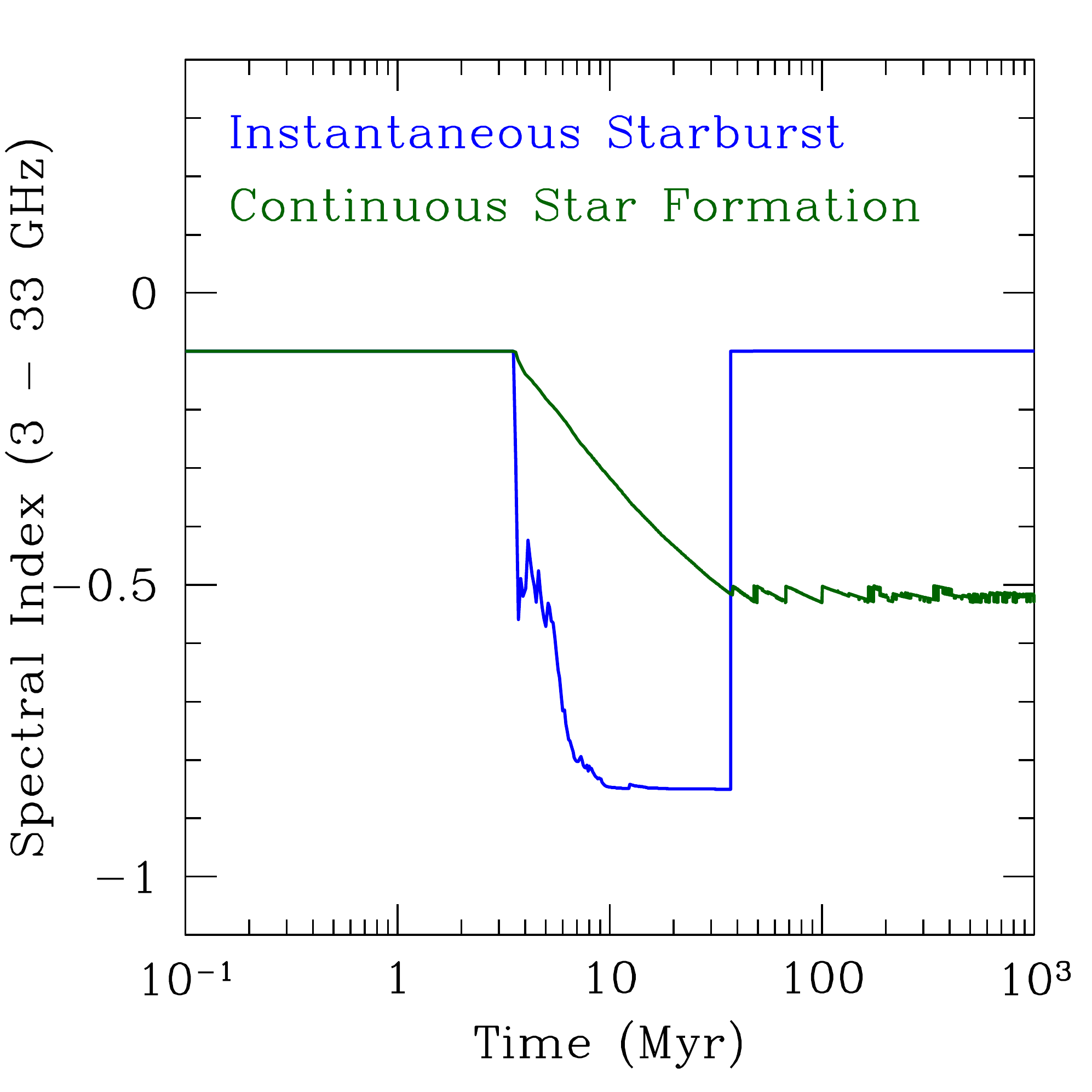}
\includegraphics[scale=1]{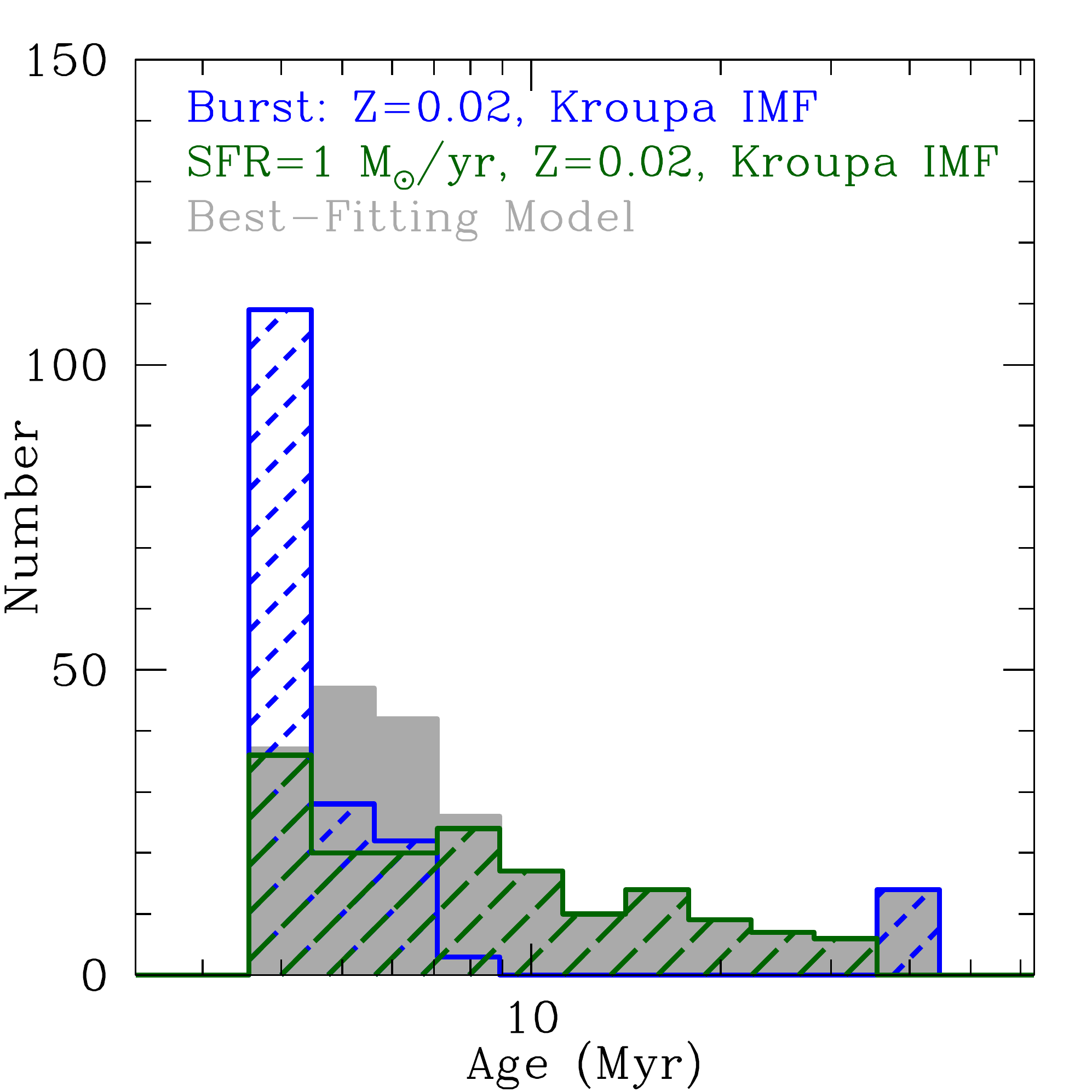}}
\vskip -5pt
\caption{\footnotesize 
Left panel: The evolution of the 3-33 GHz spectral slope in SB99 models of both an instantaneous burst and continuous SFH, using standard Kroupa IMF, and solar metallicities. We perform a $\chi^{2}$-minimization of these models to the observed $3-33$\,GHz spectral index of each region. Right panel: The distribution of model ages for both types of SFH (blue and green) and the best-fitting model in each case (grey). It is clear that there exists two populations of regions: Those younger than $t \sim 10$ Myr, which are best-modeled by an instantaneous burst, and those older than $t \sim 10$ Myr, which are best-modeled by a continuous SFH.}
\end{figure*}

\subsection{Likely Background Galaxies}
For the 10 likely background sources (BG) identified (i.e. sources with no obvious 8\,$\mu$m counterpart), which have a $S/N \geq 3$ in at least two radio bands, the median $3-33$\,GHz spectral index is $-0.65 \pm 0.04$ with a median absolute deviation of 0.3. This value is significantly steeper than the average value measured for the SF regions, and indicates that these sources are primarily dominated by synchrotron emission. This result further suggests that our visual classification scheme involving both radio and near-IR imaging appears to be an effective discriminator for various types of radio sources in surveys of nearby galaxies. Finally, a cross-reference with NED suggests that none of these sources have been previously identified in the literature.

\vspace{-2mm}
\subsection{Supernova and Supernova Remnants}

In order to identify possible supernova remnants, we cross-correlated our sample of 377 regions against the Open Supernova Catalog \citep[OSC:][]{osc17}. In total we identify 6 sources as being spatially coincident (within $2\arcsec$) to an identified radio source with a $S/N \geq 3$ in at least two radio bands. The $3-33$\,GHz spectral slopes measured are uniformly distributed from $-1$ to 0.5. This scatter is likely driven by SNe/R at various stages of their evolution, and therefore a large range in the degree of energy loss of the CRs as they propagate through the ISM. For one region identified in NGC\,7331, the measured 15\,GHz flux density is larger by over an order of magnitude due to a supernova, SN2014C, which was discovered in January of 2014, between the time our 3 and 15\,GHz observations of this source were taken \citep{shiv19}.

\vspace{-2mm}
\subsection{Anomalous Microwave Emission}
%Since its initial discover in 1996, AME has been observed in a range of environments.
Anomalous Microwave Emission (AME) is a known dust-correlated component of Galactic emission that has been detected by cosmic microwave background (CMB) experiments and other radio/microwave instruments at frequencies $10-60$\,GHz since the mid-1990s \citep[see][and articles within for recent reviews]{cd18}. AME is found to be spatially correlated with far-infrared thermal dust emission, but cannot be explained by synchrotron, or free-free emission mechanisms, and is far in excess of the emission contributed by thermal dust with the power-law opacity consistent with observations at sub-mm wavelengths. The most natural explanation for AME is rotational (electric dipole) emission from ultra-small dust grains \citep[i.e., `spinning dust':][]{wce57,dl98b}. The emission forms part of the diffuse Galactic foregrounds that contaminate CMB data, which operate in the frequency range $30-300$\,GHz, and hence knowledge of the spatial structure and spectral shape can inform CMB component separation. However, spinning dust emission depends critically on the dust grain size distribution, the type of dust, and the environmental conditions (density, temperature, interstellar radiation field, etc.). Thus, precise measurements of AME can also provide a new window into the ISM, complementing other multiwavelength tracers.

A number of searches for extragalactic AME have been undertaken with WMAP and Planck data \citep[e.g., the Magellanic Clouds and NGC4945:][]{cbot10,peel11}, all of which were inconclusive. Most recently, we have identified two additional detections of AME in the SFRS sample as having anomalously high 33\,GHz-to-$24\mu$m flux ratios \citep[NGC\,6946\,E4 and NGC\,4725\,B:][]{ejm10,ejm18b}. NGC\,4725\,B in particular appears consistent with a highly-embedded ($A_{V} > 5$ mag) nascent star-forming region, in which young ($\sim 3$ Myr) massive stars are still enshrouded by their natal cocoons of gas and dust, lacking enough supernova to produce synchrotron emission.

\begin{figure}
\center
	\includegraphics[scale=0.43]{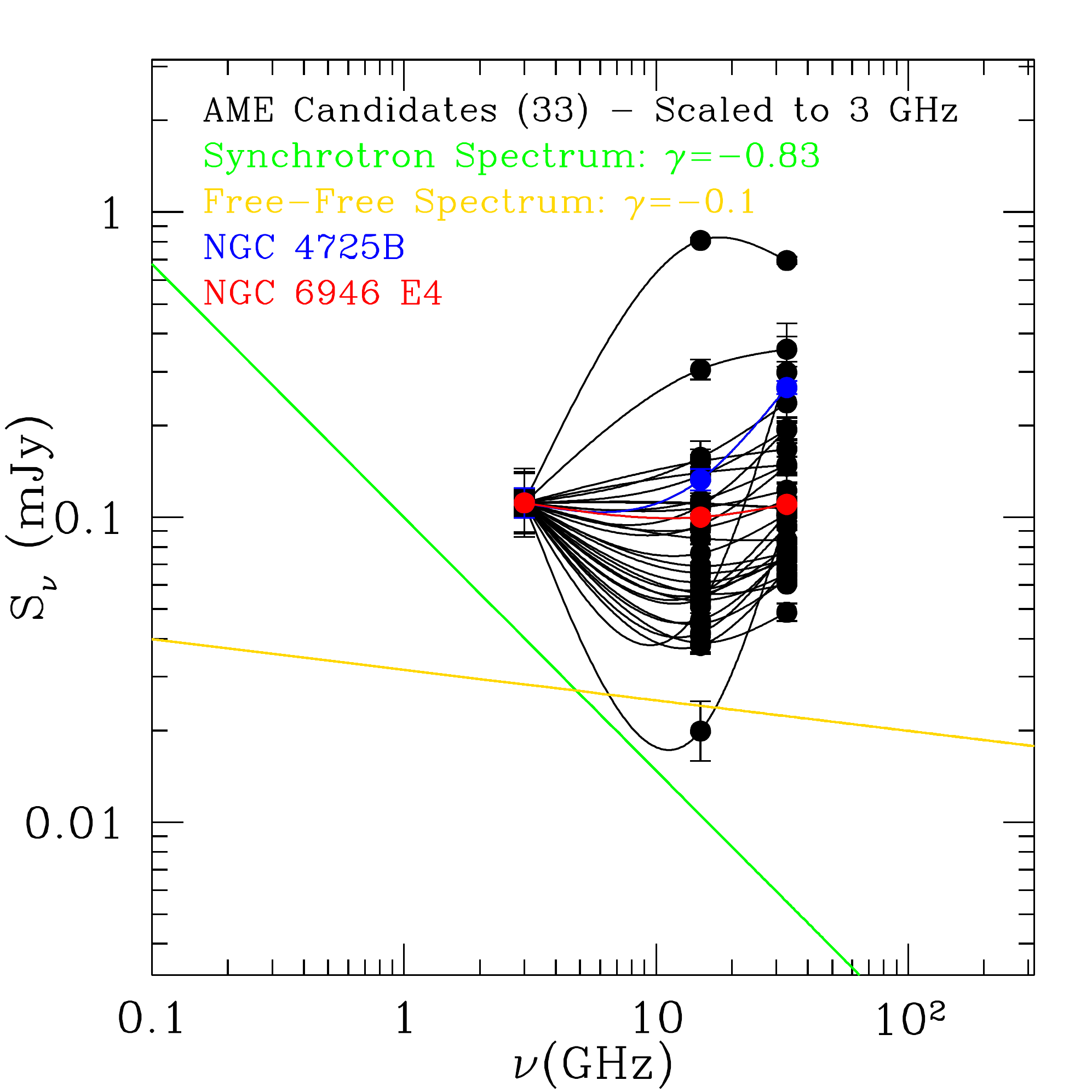}
\caption{The radio spectra of all 33 AME candidates identified in the SFRS as having $3-33$\,GHz or $15-33$\,GHz spectral indices 3$\sigma$ greater than $\alpha^{\rm T} = -0.1$. To demonstrate the diversity of our AME candidate spectra we normalize all flux density measurements to the brightest region detected at 3 GHz. We expect that, similar to NGC\,6946\,E4 (Red), the regions with decreasing $3-15$\,GHz spectral slopes contain a non-negligible amount of synchrotron emission, whereas the regions with very shallow or even rising spectra over this same frequency range (e.g. similiar to NGC\,4725\,B: Blue) will be dominated by free-free emission.}
\end{figure}

While it is possible that NGC\,4725\,B and NGC\,6946\,E4 represent the most favorable conditions for AME detection, there are likely remaining regions in the SFRS that still harbor AME at a lower level relative to the other emission components \citep{bh15}. Isolating the factors that govern the level of AME in these regions will lend insight into the physical mechanisms powering this emission as well as the nature of its carriers. In particular, we have observed that a common feature among both detections thus far is a shallow or even rising spectra from 3 - 33\,GHz, as the contribution from AME increases and eventually dominates beyond $\sim 20$\,GHz. These regions can be identified as having elevated 33\,GHz emission relative to the expected extrapolation from lower-frequency radio data using a standard two-component power law.

%33 extragalactic AME candidates in 6 SFRS galaxies
By measuring the $3-33$ and $15-33$\,GHz spectral indices, we made an initial selection of 58 extragalactic AME candidates as regions that have an 8$\mu$m counterpart, a $S/N \geq 3$ in at least 2 radio bands, a de-projected galactocentric radius of $r_{\rm G} > 250$pc, and a measured spectral slope that is $\geq 3\sigma$ above the canonical $\alpha^{\rm T} = -0.1$ value expected for free-free emission. This represents a conservative upper-limit to identify regions by assuming that the non-thermal synchrotron emission is negligible at 33\,GHz, and thus does not contribute to the measured slope from $15-33$\,GHz. This assumption is well-supported here for our full spectral analysis of over 300 star-forming regions identified in the SFRS, which show that the median thermal fraction measured at 33\,GHz on a few $\sim 100$ pc scales is $\sim 91\%$. We then visually inspected all 58 AME candidates, only retaining 33 regions where the emission was found to be compact, similar to the previously identified extragalactic AME sources. This is done to ensure that any differences in surface brightness sensitivity between our observations would not result in diffuse emission with an artificially flattened spectrum, which is unassociated with an individual source. Our final requirement removes 4 sources (NGC\,4254\,Enuc.\,1\,B, NGC\,3521\,Enuc.\,1, NGC\,4736\,N, and NGC\,7331\,D), which have very steeply rising 3-15 GHz spectral slopes, but are undetected at 33 GHz, and therefore are not confidently identified as AME candidates. 

In Figure 8 we plot the $3 - 33$\,GHz spectrum of the 33 AME candidate regions normalized to the highest measured 3\,GHz flux density. Viewed in this way we see that many of our AME candidates, similar to NGC\,4725 B, have shallow or slightly negative slopes from $3 - 15$\,GHz, and a much steeper positive slope from $15 - 33$\,GHz. However, there are some regions which look more similar to NGC\,6946 Enuc 4., which have steeper $3-15$\,GHz slopes and a less significant increase from $15-33$\,GHz. Importantly, these spectra cannot be explained by a simple combination of synchrotron (green) and free-free (yellow) emission components, suggesting that either an additional emission component peaking at $\gtrsim 15$\,GHz is required (e.g., AME), it is a (very) high-frequency GHz-peaked background galaxy, or the source is variable. A final possibility is that the free-free emission is optically-thick at 33 GHz. However, such `ultra-compact' H{\sc ii} regions have much higher radio luminosities and SFRs ($\sim40-60$ mJy) then the regions identified here \citep[e.g.,][]{dm02}.

\section{Conclusions}

We have presented 3, 15, and 33\,GHz imaging towards galaxy nuclei and extranuclear star-forming regions in the SFRS, and have identified 335 regions (286 SF, 10 BG, 6 SNe/R, and 33 AME) with $S/N \geq 3$ in at least two radio bands. Through detailed measurements of their radio spectra we have confirmed that:

\begin{enumerate}
\item The average local background contribution to the measured 3, 15, and 33\,GHz flux densities on $\sim 100$ pc scales is $\sim 4-6\%$. This is significantly smaller than the $15-40\%$ found for the sample of regions studied at $25\arcsec$ ($\sim 1$kpc) scales with the GBT \citep{ejm11b}.

\item On $\sim$100\,pc scales, the median thermal fraction at 33\,GHz of all regions identified as non-background galaxies is $92 \pm 0.8\%$ with a median absolute deviation of $11\%$. Limiting our analysis to extranuclear ($r_{\rm G} > 250$pc) SF regions, we measure a median thermal fraction of $93 \pm 0.8\%$ with a median absolute deviation of $10\%$. Further, we find that on 7\arcsec~scales the median thermal fraction is $94 \pm 0.8\%$, and thus the thermal fraction remains $\geq 90\%$ up to $\sim 500$ pc scales. 
%Finally, we find that isolating extranuclear SF regions at both native and 7\arcsec resolution results in a higher thermal fraction at 33\,GHz and a smaller median absolute deviation in the overall distribution.

\item We have confirmed through MCMC analysis that we do not introduce systematic biases when interpreting the results of the $\chi^{2}$-minimization of a two-component power-law model to fit the observed radio spectrum from $3 - 33$\,GHz, and that this model can adequately separate the thermal free-free and nonthermal synchrotron emission components over a realistic range of input values.

\item We find a systematic increase in the scatter of the measured spectral indices and thermal fractions as the de-projected galactocentric radius approaches the nucleus. This trend is reflective of the ongoing star-formation activity occurring in centers of these galaxies, and results in a larger contribution of diffuse nonthermal emission.

\item We have identified a sample of 33 sources whose rising $15-33$\,GHz emission may be due to anomalous microwave emission. Follow-up observations at high ($\geq 40$\,GHz) frequencies will be necessary to confirm these sources as discrete regions of extragalactic AME.

\end{enumerate}

\acknowledgements
S.T.L. was supported by the NRAO Grote Reber Dissertation Fellowship. The National Radio Astronomy Observatory is a facility of the National Science Foundation operated under cooperative agreement by Associated Universities, Inc. 
%A.S.E., and Y.S. were supported by NSF grant AST 1816838. 
%G.C.P. acknowledges support from the University of Florida. 
%A.S.E. was also supported by the Taiwan, R.O.C. Ministry of Science and Technology grant MoST 102-2119-M-001-MY3. T.D-S. acknowledges support from ALMA-CONICYT project 31130005 and FONDECYT regular project 1151239.
%Portions of this work were performed at the Aspen Center for Physics, which is supported by National Science Foundation grant PHY-1066293. This work was partially supported by a grant from the Simons Foundation. 
This research has made use of the NASA/IPAC Extragalactic Database (NED) which is operated by the Jet Propulsion Laboratory, California Institute of Technology, under contract with the National Aeronautics and Space Administration.
%This research has made use of the NASA/ IPAC Infrared Science Archive, which is operated by the Jet Propulsion Laboratory, California Institute of Technology, under contract with the National Aeronautics and Space Administration.

\bibliography{master_ref}

\appendix
\startlongtable
\begin{deluxetable*}{l|cccc}
\tablecolumns{5}
\tablecaption{Ancillary Source Photometry at 7\arcsec~Angular Resolution \label{tbl-A2}}
\tablewidth{0pt}
\tablehead{
\colhead{Source ID} & \colhead{$f_{\nu} (1528$\,\AA)}  & \colhead{$f_{\nu} (2271$\,\AA)}  & \colhead{$f_{\rm H\alpha}/10^{-13}$} & \colhead{$f_{\nu} (24\,\micron)$}\\
\colhead{} & \colhead{($\mu$Jy)}  & \colhead{($\mu$Jy)}  & \colhead{(erg\,s$^{-1}$\,cm$^{-2}$)} & \colhead{(mJy)}
}
\startdata
\cutinhead{Star-Forming Regions}
                            NGC\,0337\,a   & $   76.38\pm 11.46$  & $  163.59\pm 24.54$  &   $  1.47\pm  0.29$  &            $  45.73\pm  2.29$  \\
                            NGC\,0337\,b   & $  104.70\pm 15.71$  & $  185.04\pm 27.76$  &   $  1.83\pm  0.37$  &            $  81.65\pm  4.08$  \\
                            NGC\,0337\,c   & $   36.93\pm  5.54$  & $   70.52\pm 10.58$  &   $  0.57\pm  0.11$  &            $   8.40\pm  0.42$  \\
                            NGC\,0337\,d   & $   47.19\pm  7.08$  & $   87.63\pm 13.15$  &   $  1.18\pm  0.24$  &            $  22.25\pm  1.11$  \\
                     NGC\,0628\,Enuc.\,4   & $   67.64\pm 10.15$  & $   81.23\pm 12.19$  &   $  0.53\pm  0.11$  &            $   9.87\pm  0.50$  \\
                     NGC\,0628\,Enuc.\,2   & $   70.55\pm 10.58$  & $   95.46\pm 14.32$  &   $  0.40\pm  0.08$  &            $  40.37\pm  2.02$  \\
                     NGC\,0628\,Enuc.\,3   & $  233.35\pm 35.00$  & $  282.27\pm 42.34$  &   $  0.78\pm  0.16$  &            $  28.94\pm  1.45$  \\
                               NGC\,0628   & $   16.96\pm  2.54$  & $   41.26\pm  6.19$  &   $  0.06\pm  0.01$  &            $   2.41\pm  0.14$  \\
\enddata
\tablecomments{$^{\dagger}$\,Photometry unreliable and not reported due to saturation by the nucleus in the 24\,$\mu$m image.}
\end{deluxetable*}

\startlongtable
\begin{deluxetable*}{l|ccc|ccc}
\tablecolumns{7}
\tablecaption{Local Background Measurements \label{tbl-A1}}
\tablewidth{0pt}
\tablehead{
\colhead{Source ID}  & \colhead{Back$_\mathrm{3\,GHz}$} & \colhead{Back$_\mathrm{15\,GHz}$} & \colhead{Back$_\mathrm{33\,GHz}$} & \colhead{f$_{Back}^\mathrm{3\,GHz}$} & \colhead{f$_{Back}^\mathrm{15\,GHz}$} & \colhead{f$_{Back}^\mathrm{33\,GHz}$}  \\
\colhead{} & \colhead{(mJy)}  & \colhead{(mJy)}  & \colhead{(mJy)} & \colhead{($^{\ddagger}$)} & \colhead{($^{\ddagger}$)} &  \colhead{($^{\ddagger}$)} 
}
\startdata
\cutinhead{Star-Forming Regions}
                            NGC\,0337\,D   &             0.207  &             0.099  &             0.019  &             0.174  &             0.130  &             0.062  \\
                            NGC\,0337\,A   &             0.046  &             0.070  &             0.024  &             0.046  &             0.100  &             0.085  \\
                            NGC\,0337\,B   &             0.133  &             0.059  &            -0.022  &             0.027  &             0.022  &             0.012  \\
                            NGC\,0337\,C   &             0.026  &             0.019  &             0.005  &             0.087  &             0.064  &             0.024  \\
                            NGC\,0337\,G   &             0.030  &             0.013  &            -0.005  &             0.082  &             0.068  &             0.032  \\
                            NGC\,0337\,E   &             0.026  &             0.027  &             0.010  &             0.096  &             0.099  &             0.034  \\
                            NGC\,0337\,F   &             0.021  &             0.038  &             0.023  &             0.075  &             0.195  &             0.104  \\
                     NGC\,0628\,Enuc.\,4   &            -0.004  &             0.013  &             0.013  &             0.015  &             0.049  &             0.057  \\
\enddata
\tablecomments{$^{\dagger}$ The aperture used for this star-forming region contained multiple smaller individual regions.}
\tablecomments{$^{\ddagger}$ The fractional contribution of the absolute value of the local background measurement to the aperture flux density.}
\end{deluxetable*}

\begin{figure*}
\center
	\includegraphics[scale=0.26]{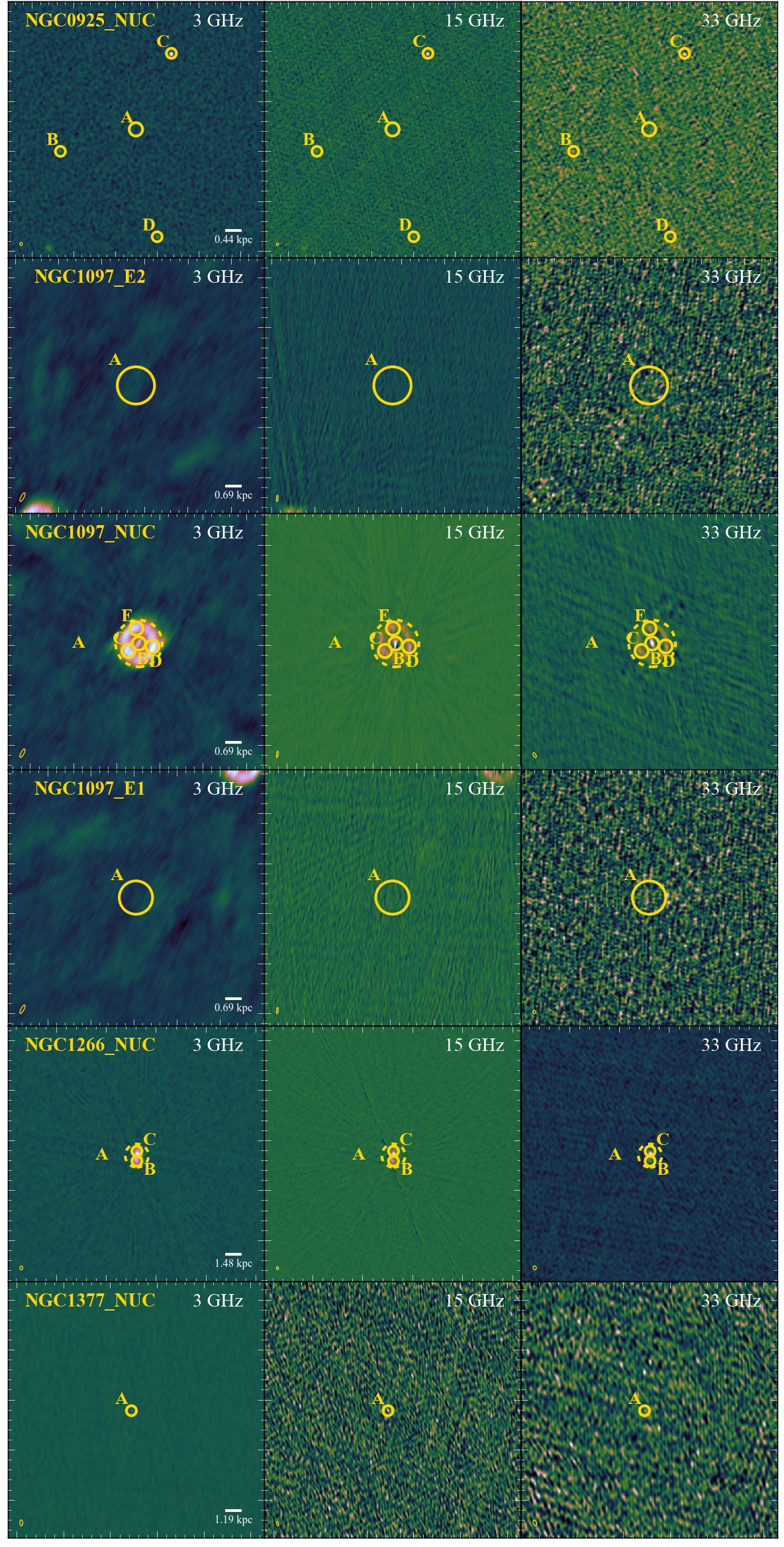}
\caption{See description in Figure 1.}
\end{figure*}

\begin{figure*}
\center
	\includegraphics[scale=0.26]{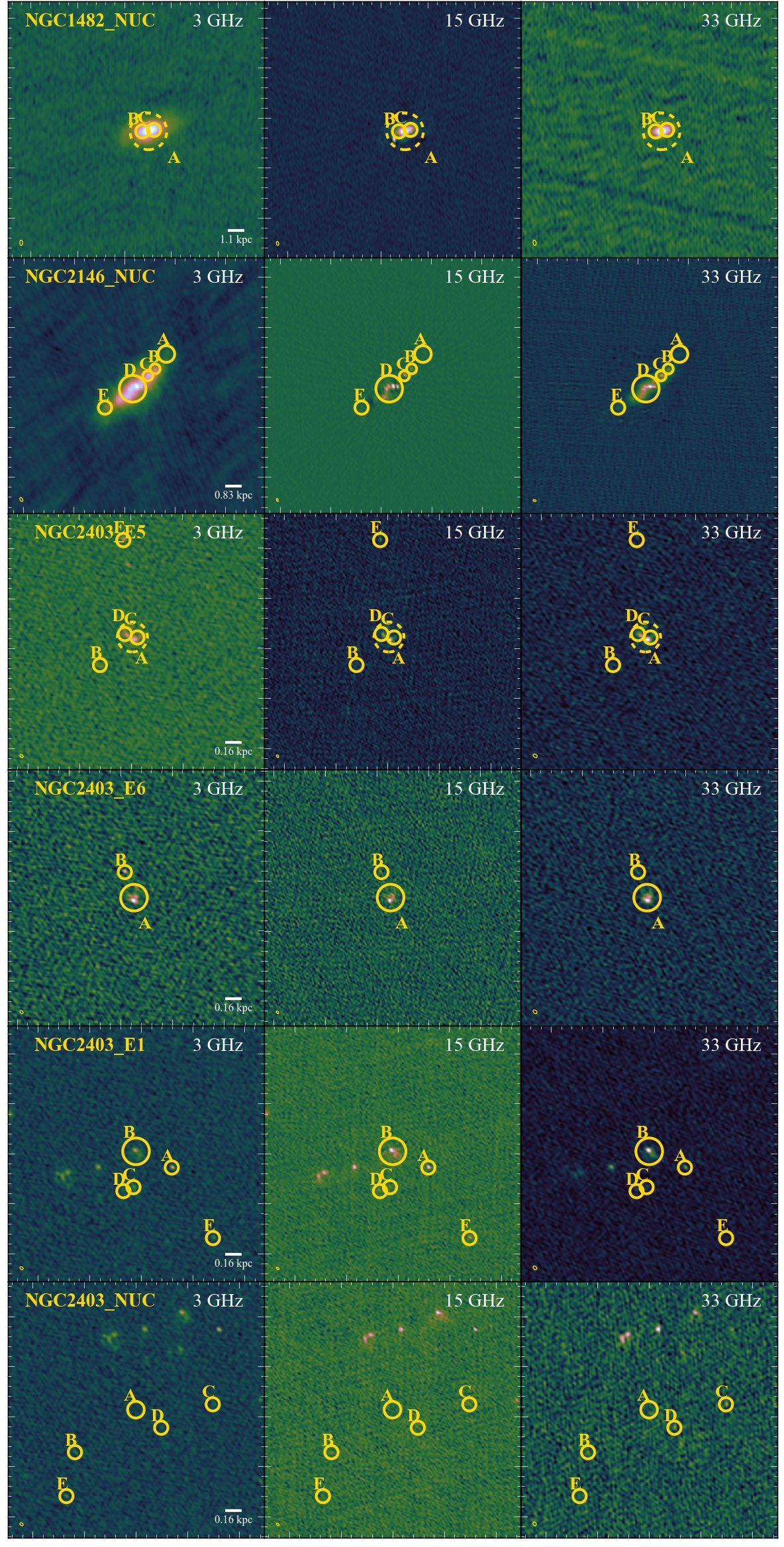}
\caption{See description in Figure 1.}
\end{figure*}

\begin{figure*}
\center
	\includegraphics[scale=0.26]{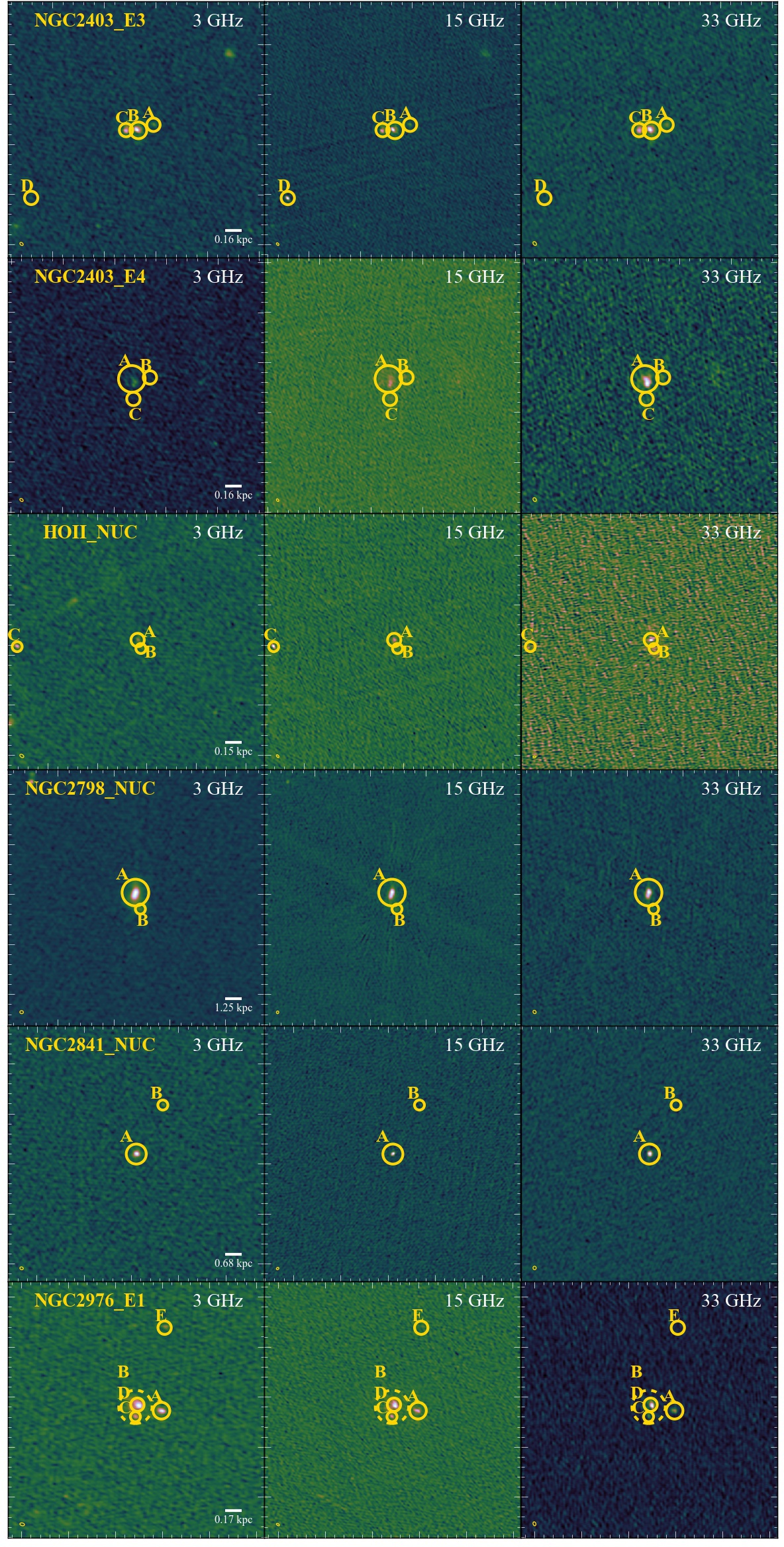}
\caption{See description in Figure 1.}
\end{figure*}

\begin{figure*}
\center
	\includegraphics[scale=0.26]{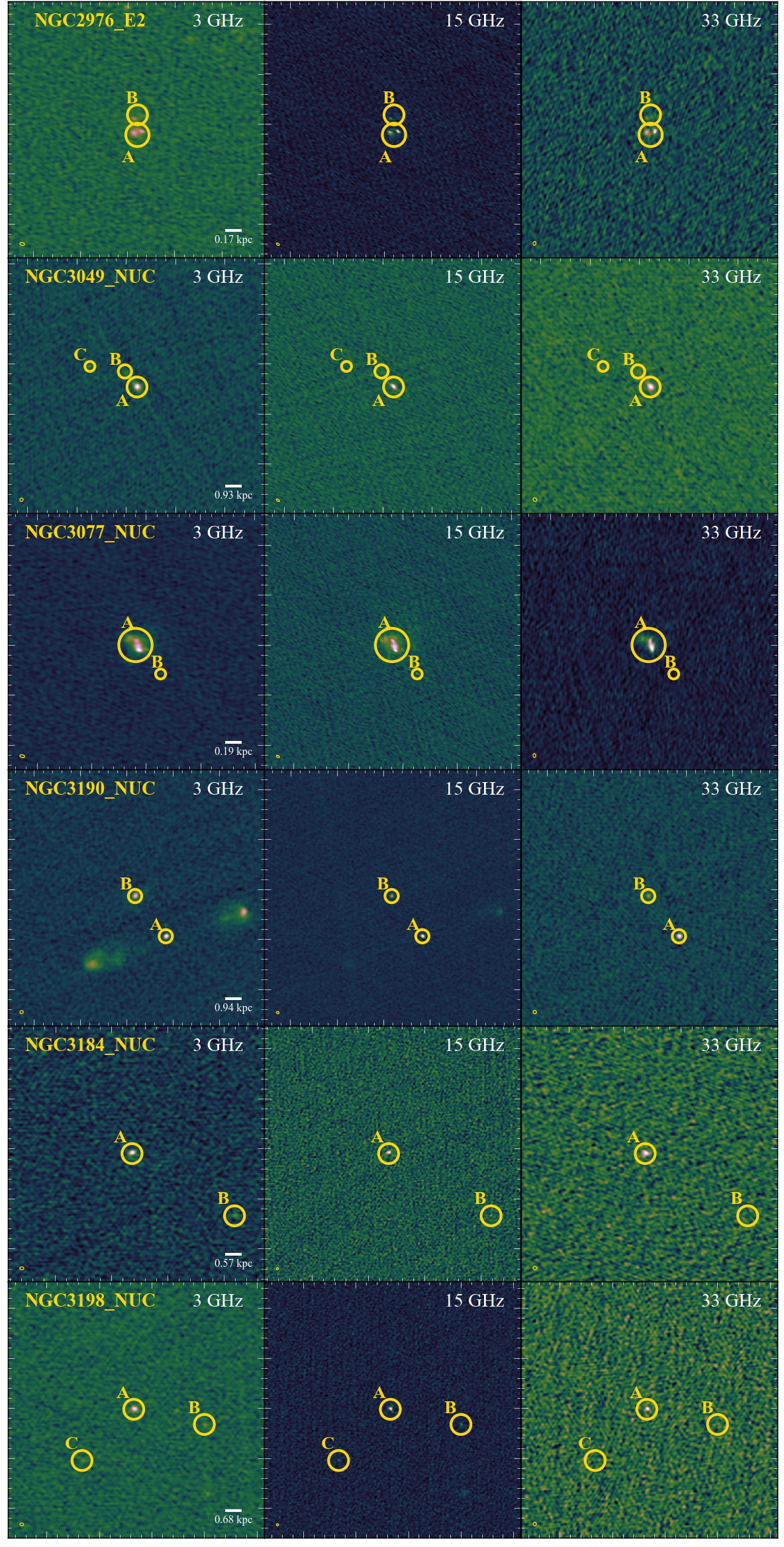}
\caption{See description in Figure 1.}
\end{figure*}

\begin{figure*}
\center
	\includegraphics[scale=0.26]{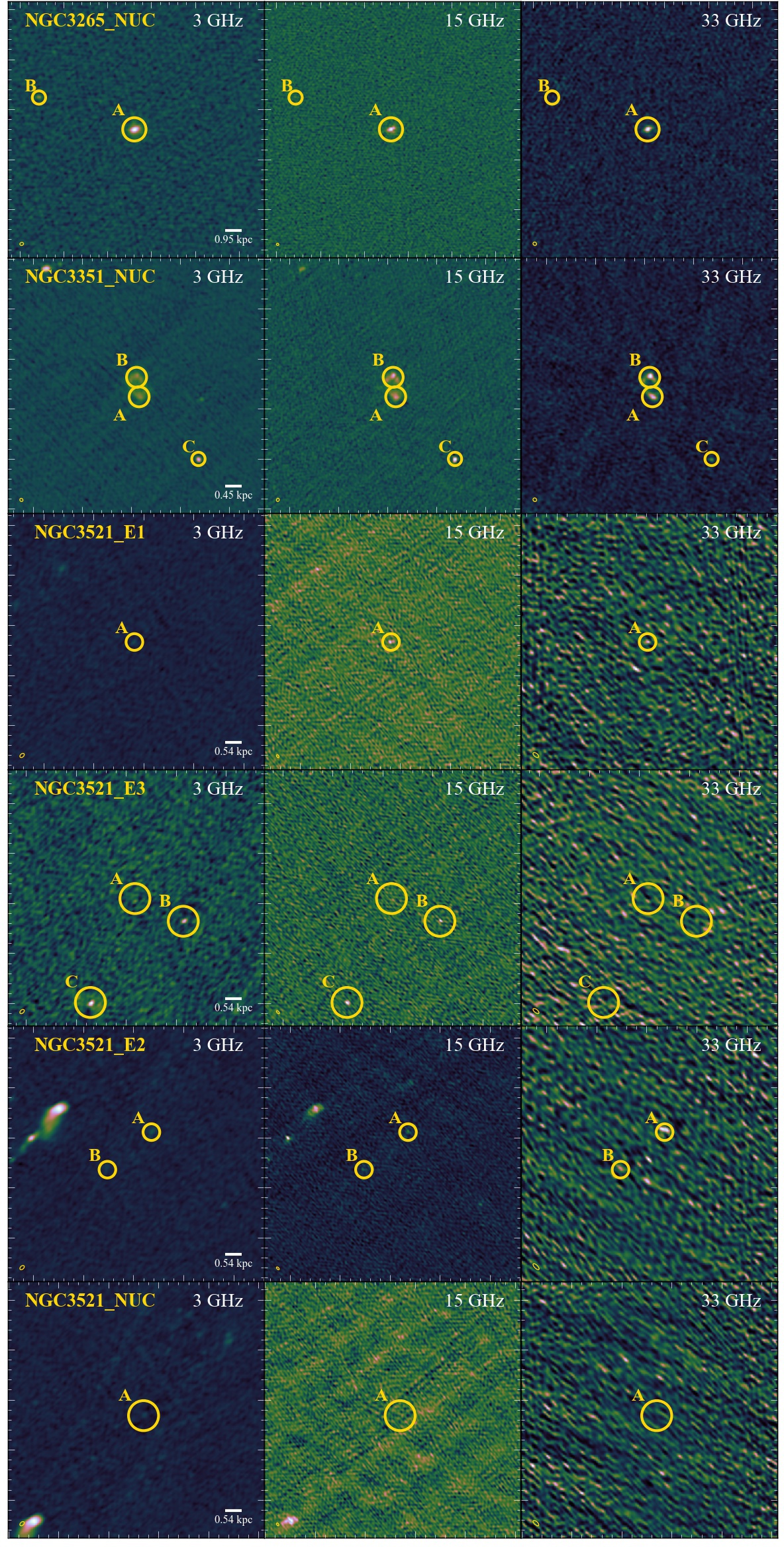}
\caption{See description in Figure 1.}
\end{figure*}

\begin{figure*}
\center
	\includegraphics[scale=0.26]{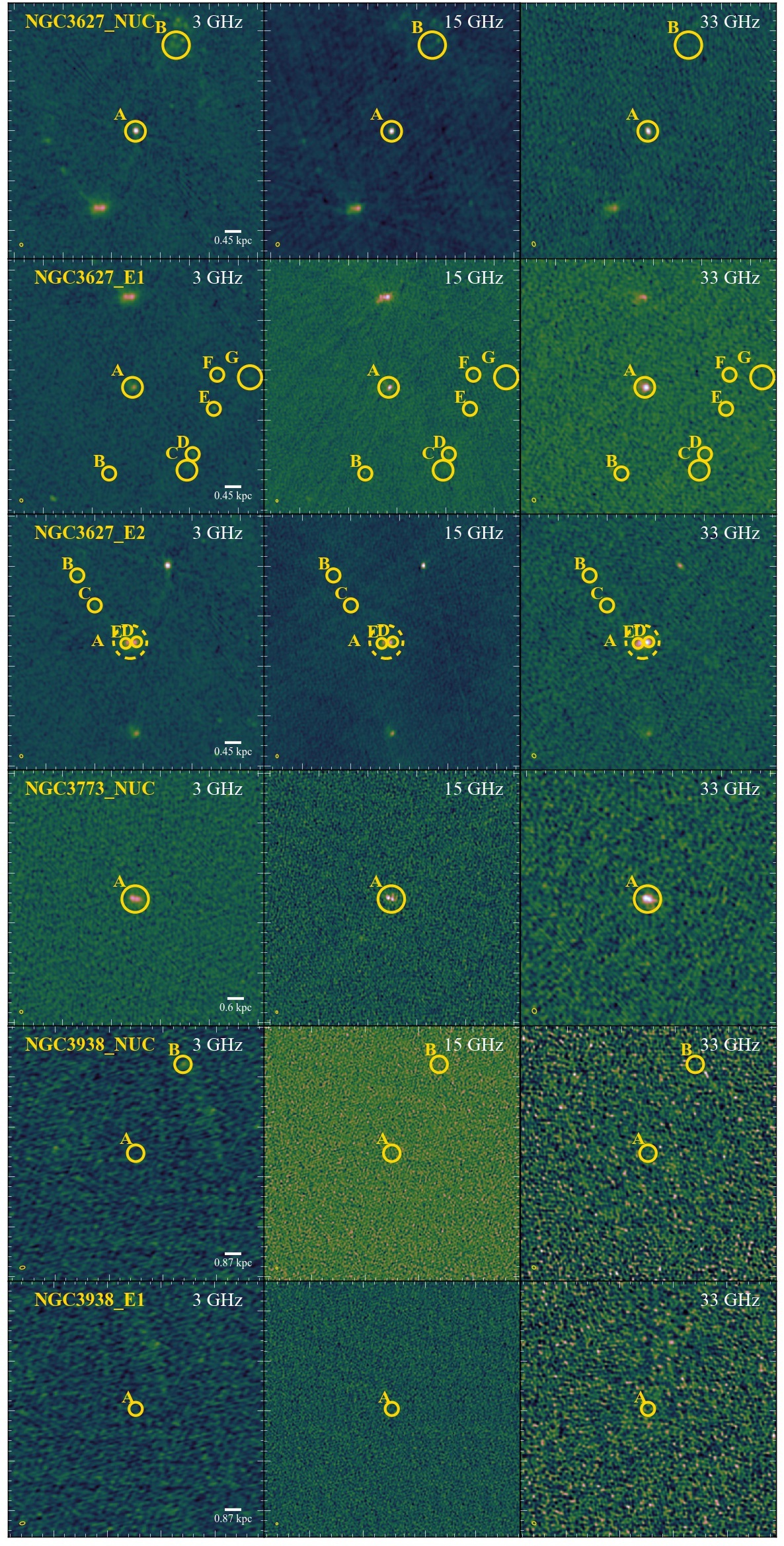}
\caption{See description in Figure 1.}
\end{figure*}

\begin{figure*}
\center
	\includegraphics[scale=0.26]{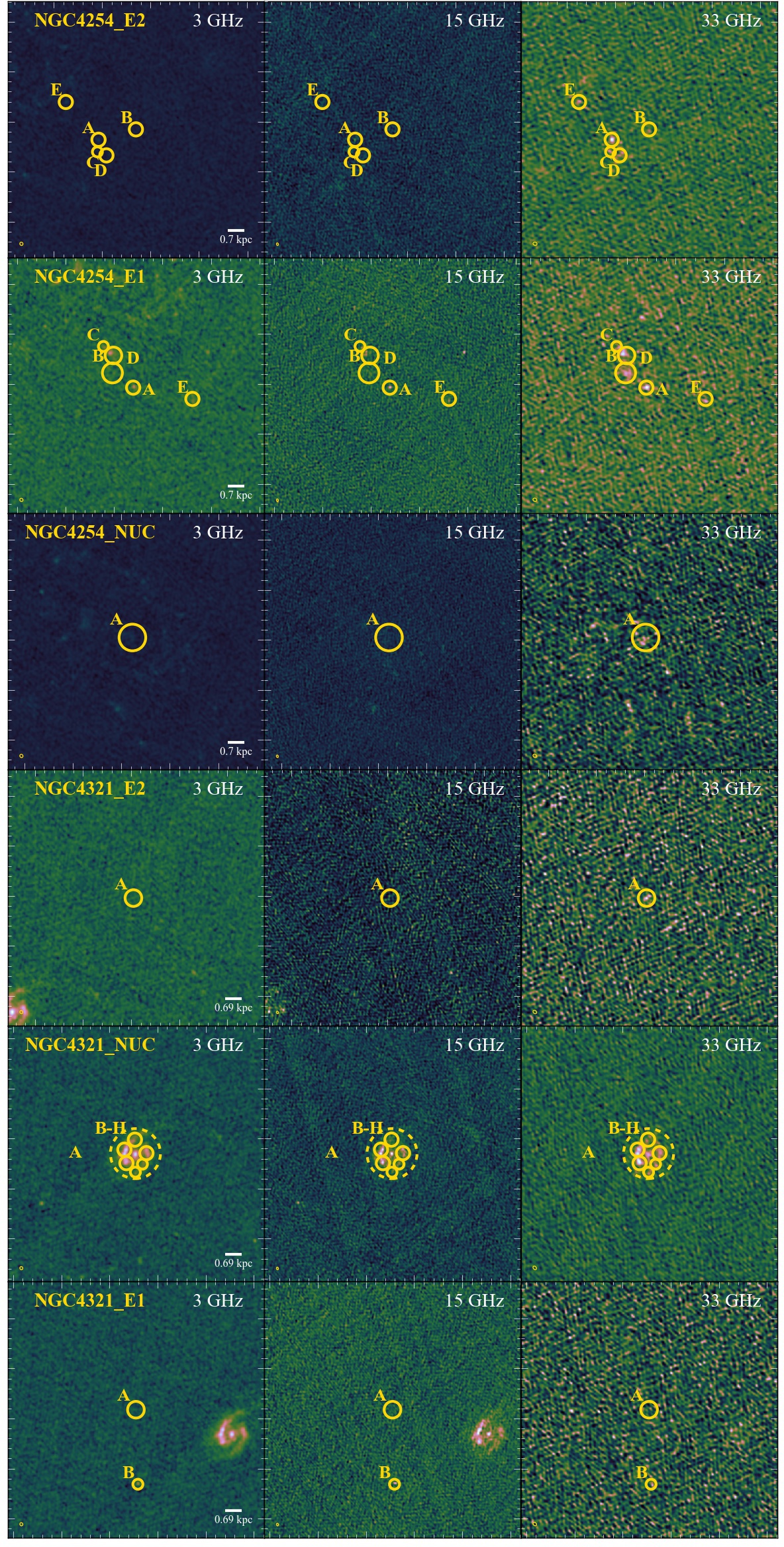}
\caption{See description in Figure 1.}
\end{figure*}

\begin{figure*}
\center
	\includegraphics[scale=0.26]{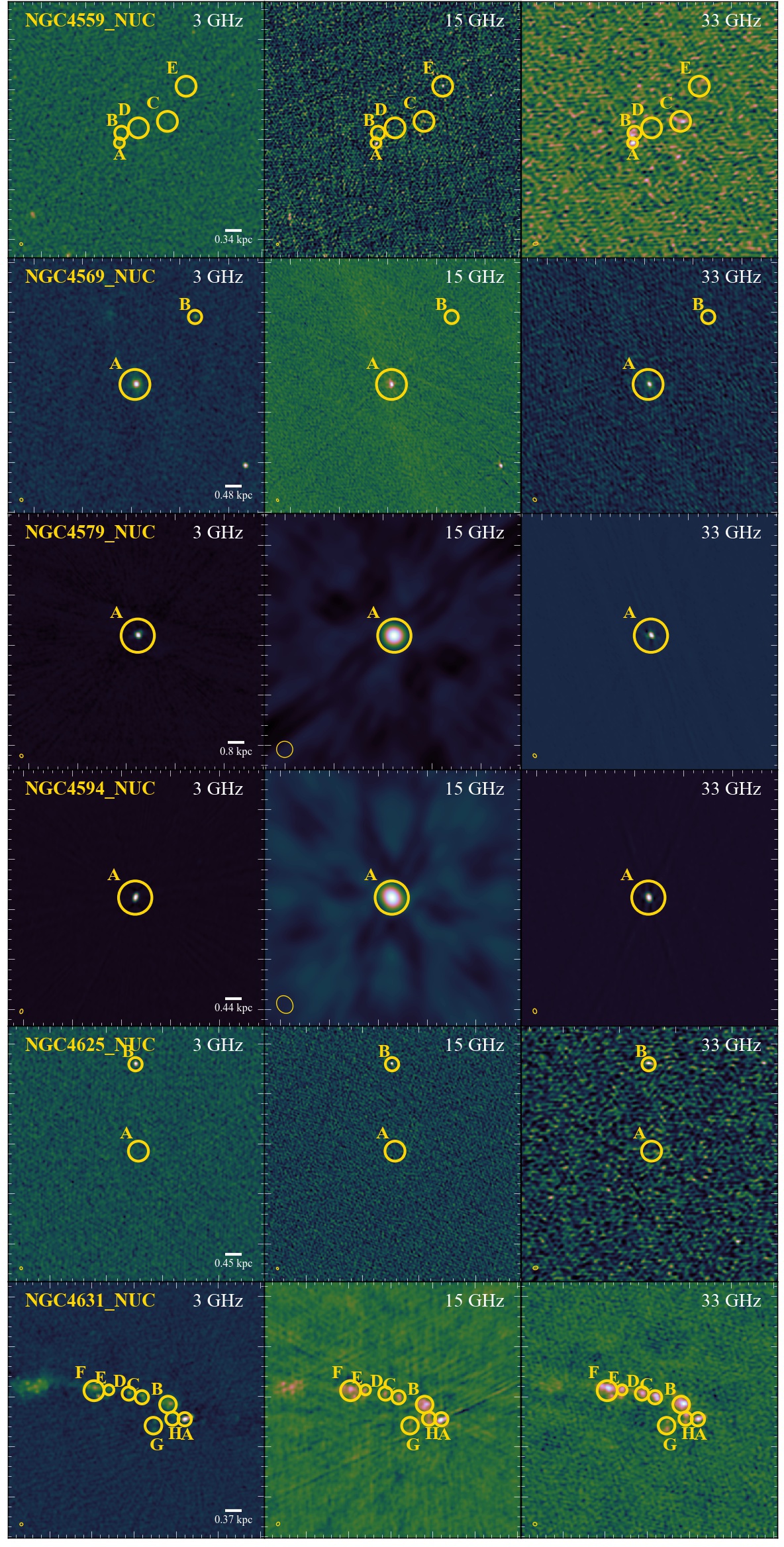}
\caption{See description in Figure 1.}
\end{figure*}

\begin{figure*}
\center
	\includegraphics[scale=0.26]{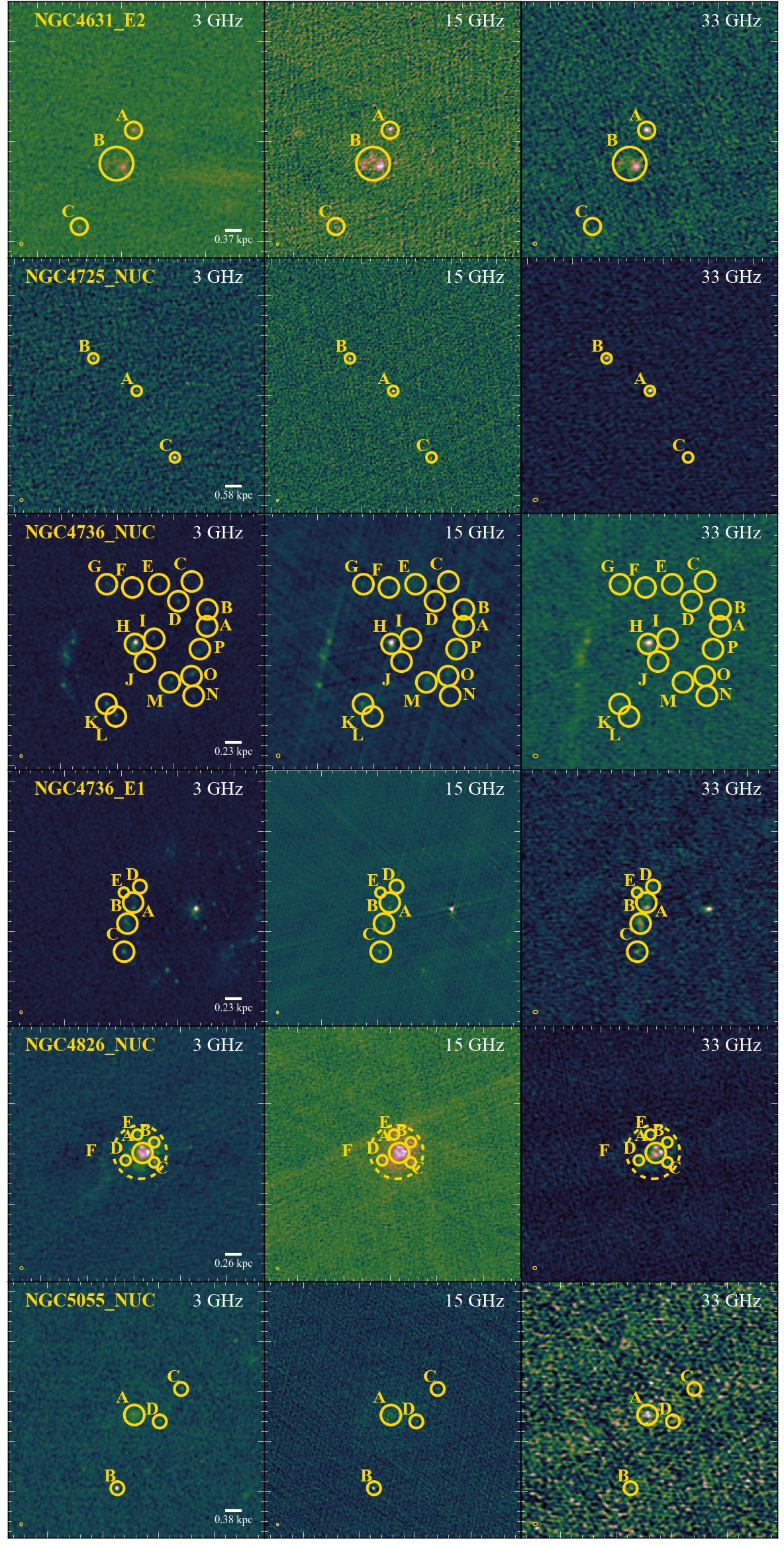}
\caption{See description in Figure 1.}
\end{figure*}

\begin{figure*}
\center
	\includegraphics[scale=0.26]{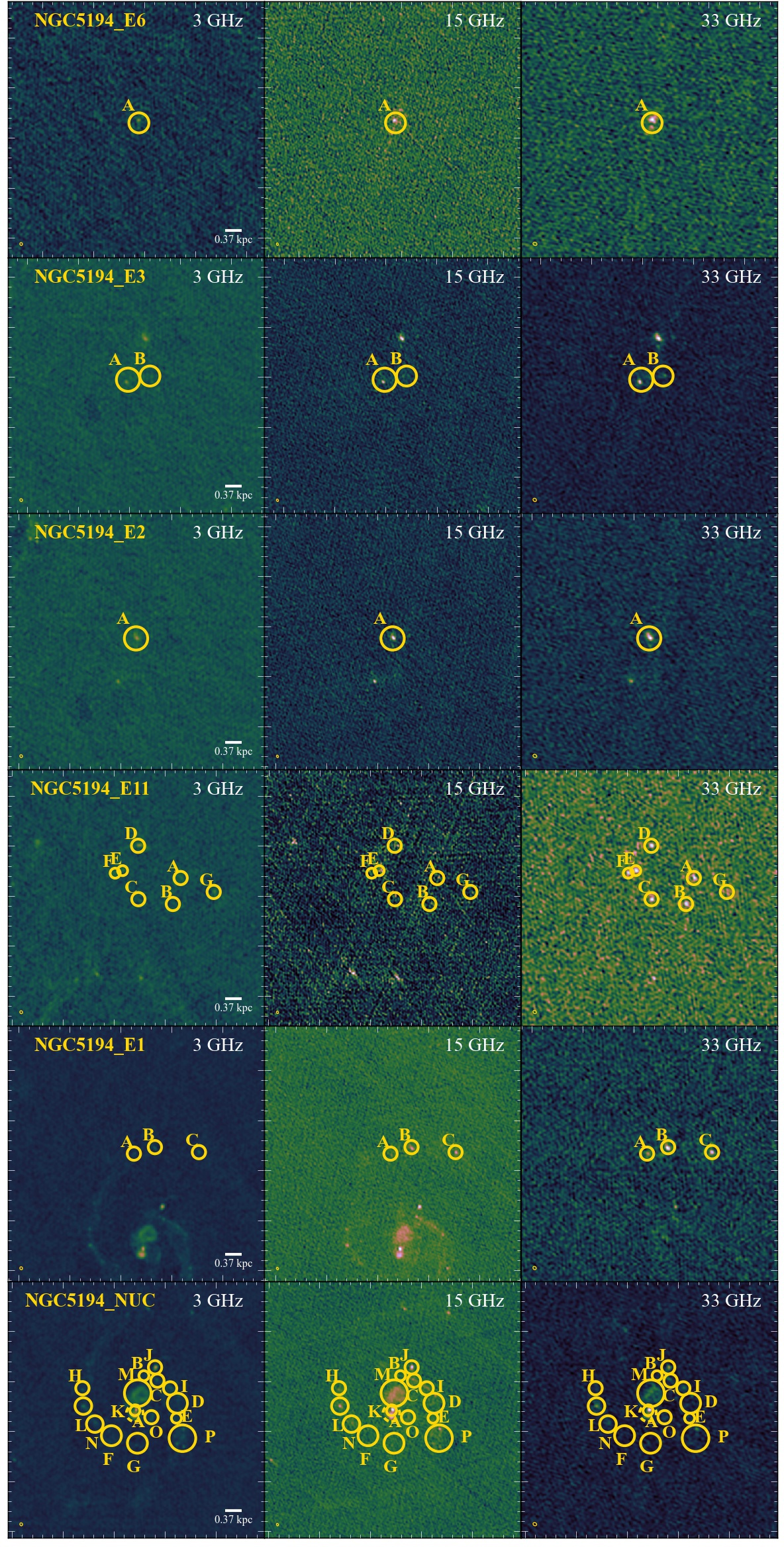}
\caption{See description in Figure 1.}
\end{figure*}

\begin{figure*}
\center
	\includegraphics[scale=0.26]{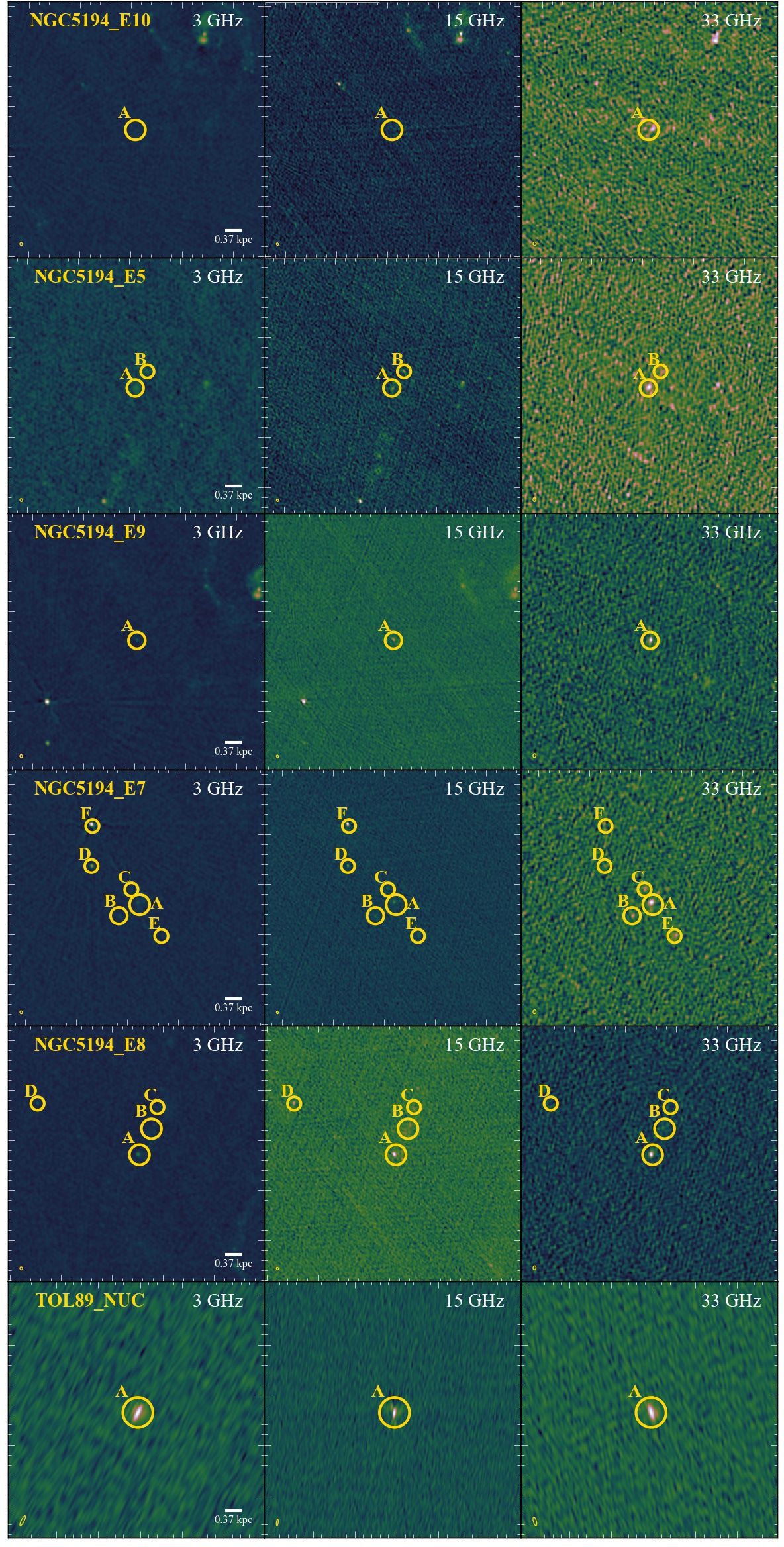}
\caption{See description in Figure 1.}
\end{figure*}

\begin{figure*}
\center
	\includegraphics[scale=0.26]{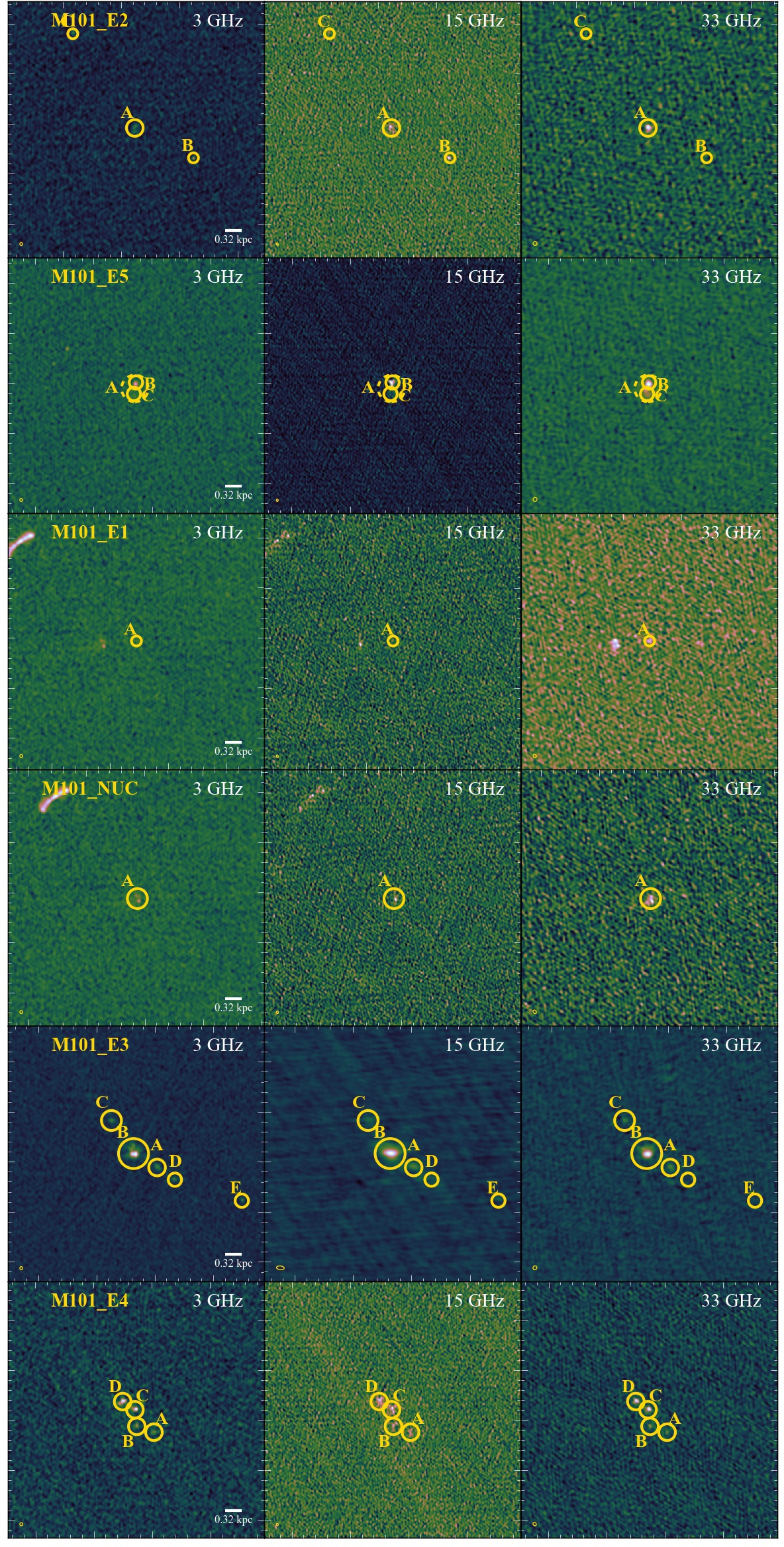}
\caption{See description in Figure 1.}
\end{figure*}

\begin{figure*}
\center
	\includegraphics[scale=0.26]{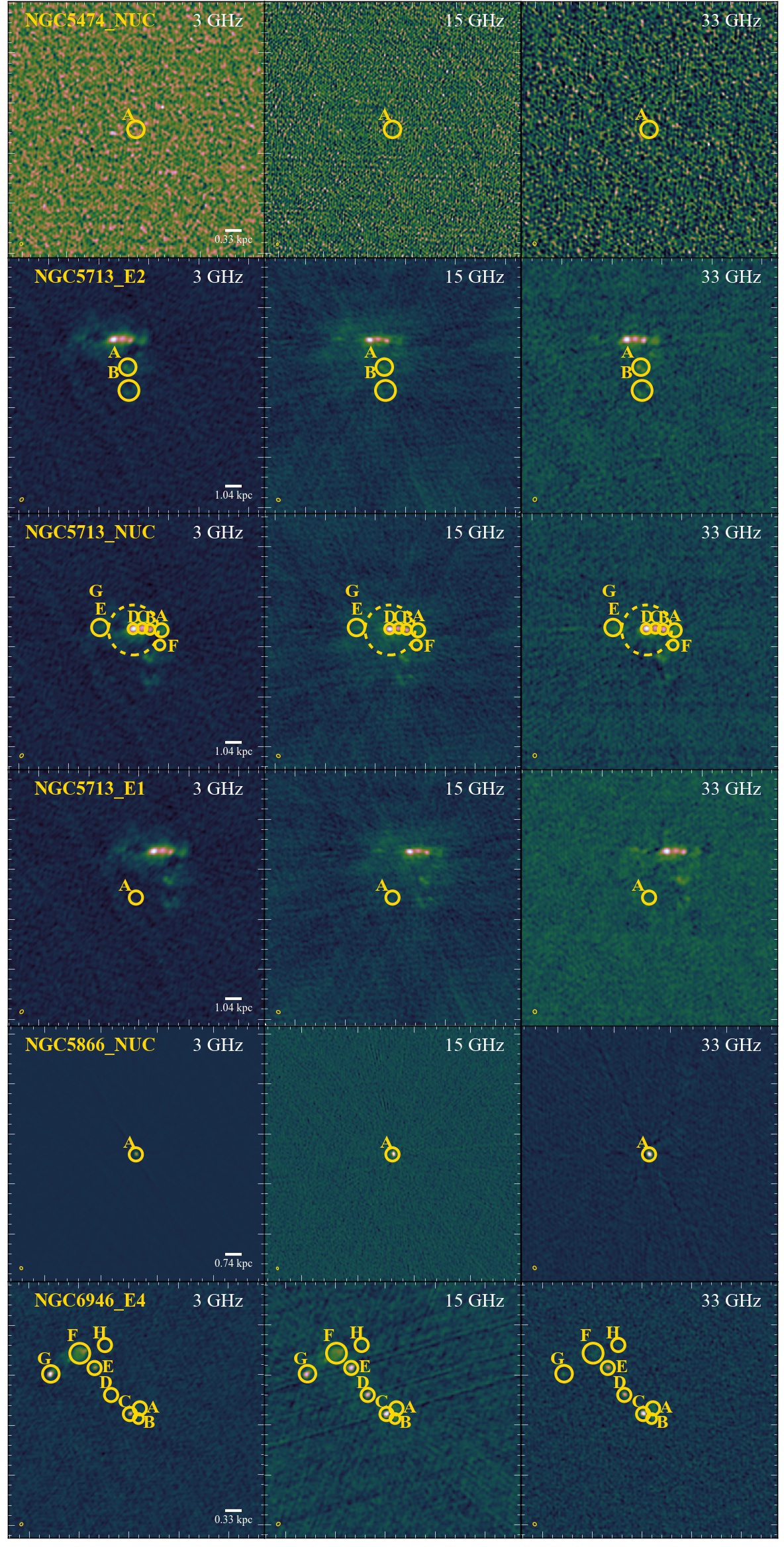}
\caption{See description in Figure 1.}
\end{figure*}

\begin{figure*}
\center
	\includegraphics[scale=0.26]{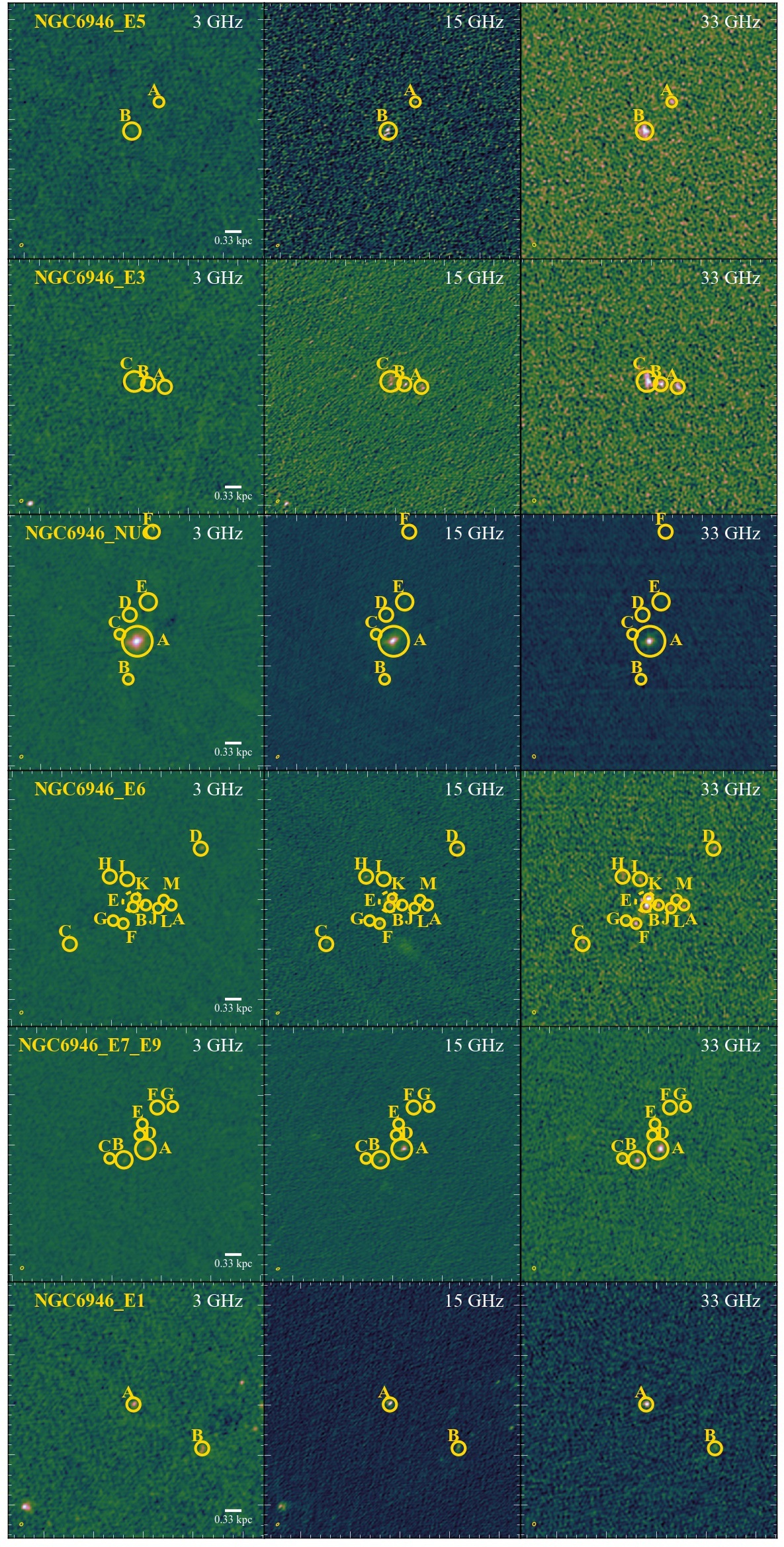}
\caption{See description in Figure 1.}
\end{figure*}

\begin{figure*}
\center
	\includegraphics[scale=0.26]{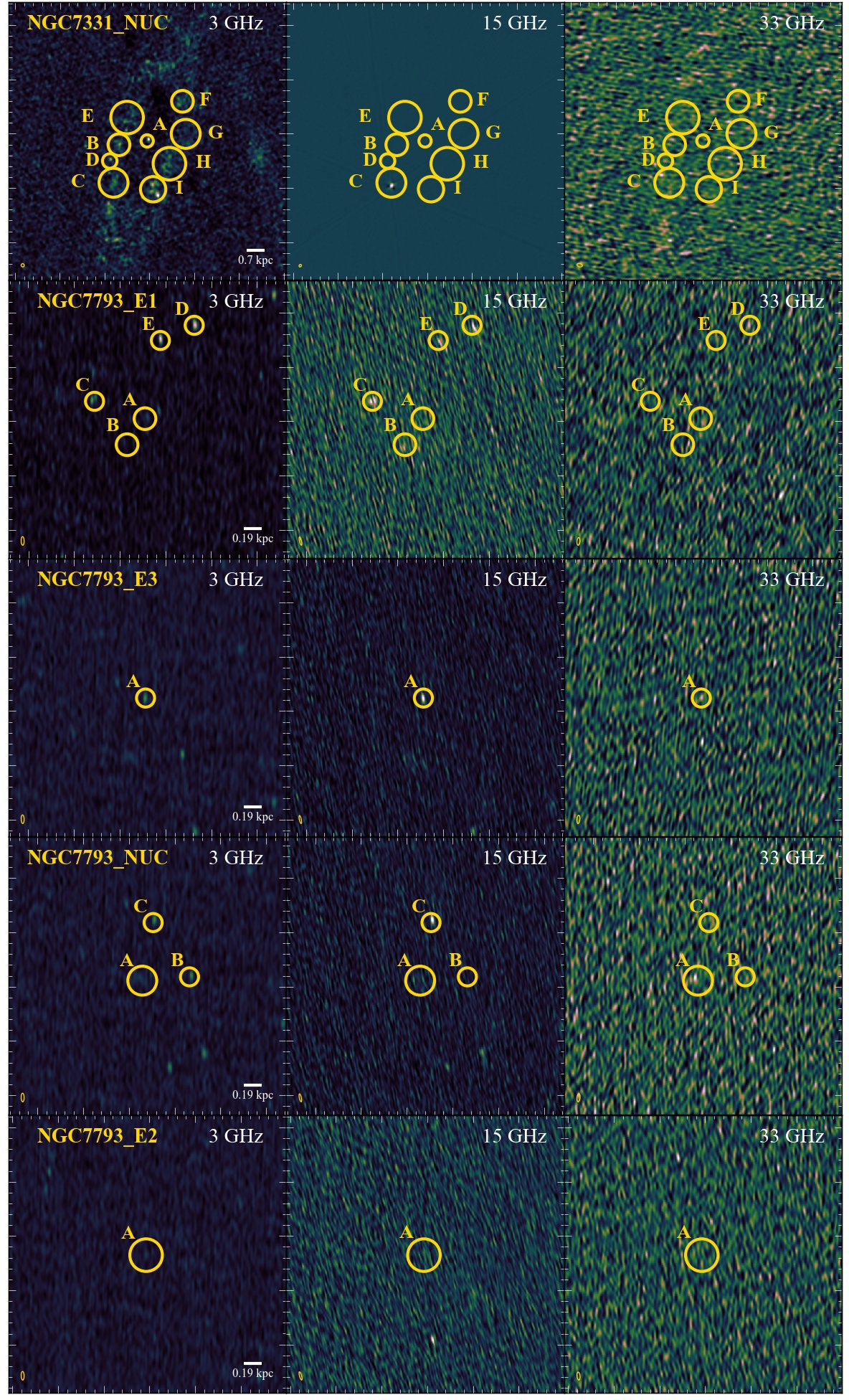}
\caption{See description in Figure 1.}
\end{figure*}

\end{document}